\documentclass{emulateapj}
\RequirePackage{lineno}
\usepackage{amssymb}
\usepackage{amsmath}
\usepackage{mathrsfs}
\usepackage{natbib}
\usepackage[colorlinks, linkcolor=blue, urlcolor=blue, citecolor=blue]{hyperref}
\usepackage{array}
\usepackage{float}
\usepackage{graphicx}
\usepackage{subfigure}
\usepackage{color}
\usepackage[all]{hypcap}
\usepackage[toc,page]{appendix}
\usepackage{multirow}

\def\beq{\begin{equation}}
\def\enq{\end{equation}}
\def\bea{\begin{eqnarray}}
\def\ena{\end{eqnarray}}

\def\apjl{ApJL\,}

\begin{document}

\title{Afterglows and Kilonovae Associated with Nearby Low-Luminosity Short-Duration Gamma-Ray Bursts: Application to GW170817/GRB170817A}
\author{Di Xiao\altaffilmark{1,2}, Liang-Duan Liu\altaffilmark{1,2}, Zi-Gao Dai\altaffilmark{1,2} and Xue-Feng Wu\altaffilmark{3,4}}
\affil{\altaffilmark{1}School of Astronomy and Space Science, Nanjing University, Nanjing 210093, China; dxiao@nju.edu.cn, dzg@nju.edu.cn}
\affil{\altaffilmark{2}Key Laboratory of Modern Astronomy and Astrophysics (Nanjing University), Ministry of Education, China}
\affil{\altaffilmark{3}Purple Mountain Observatory, Chinese Academy of Sciences, Nanjing 210008, China}
\affil{\altaffilmark{4}Joint Center for Particle, Nuclear Physics and Cosmology, Nanjing
University-Purple Mountain Observatory, Nanjing 210008, China}

\begin{abstract}
Very recently, the gravitational wave (GW) event GW170817 was discovered to be associated with the short gamma-ray burst (GRB) 170817A. Multi-wavelength follow-up observations were carried out, and X-ray, optical and radio counterparts to GW170817 were detected. The observations undoubtedly indicate that GRB170817A originates from a binary neutron star (BNS) merger. However, the GRB falls into the low-luminosity class which could have a higher statistical occurrence rate and detection probability than the normal (high-luminosity) class. This implies a possibility that GRB170817A is intrinsically powerful but we are off-axis and only observe its side emission. In this paper, we provide a timely modeling of the multi-wavelength afterglow emission from this GRB and the associated kilonova signal from the merger ejecta, under the assumption of a structured jet, a two-component jet, and an intrinsically less-energetic quasi-isotropic fireball respectively. Comparing the afterglow properties with the multi-wavelength follow-up observations, we can distinguish between these three models. Furthermore, a few model parameters (e.g., the ejecta mass and velocity) can be constrained.
\end{abstract}

\keywords{gamma-ray burst: general -- radiation mechanisms: non-thermal -- gravitational waves}

\section{Introduction}
Time domain astronomy has entered a new era since the monumental discovery of gravitational waves (GWs) by the advanced LIGO/Virgo observatories in the last two years \citep{abb16a,abb16b,abb17a,abb17b}. Since then, searching for electromagnetic (EM) counterparts to GWs has become a very urgent issue in this field. Four confirmed detections GW150914, GW151226, GW170104 and GW170814 are believed to originate from binary black hole (BBH) mergers with dozens of solar masses \citep{abb16a,abb16b,abb17a,abb17b}. However, usually we would not expect any EM counterpart from BBH mergers except for in the following specific situations \citep{con16, loeb16, per16, yam16, zhang16, demi17}. Differing with BBH mergers, binary neutron star (BNS) mergers are expected to generate several EM signals, such as short gamma-ray burst (GRB) jet emission \citep[e.g.][]{fab06, nak07, gia13, ber14, ruiz16, kath17}, cocoon prompt emission \citep{got17, laz17a, laz17b, nak17}, jet/cocoon afterglows \citep[e.g.][]{got17, lamb17, laz17a, nak17}, and kilonovae \citep[also referred to as ``macronovae'',][]{li98, kul05, met10, met12, kas13, hot15, got17, nak17}. A late-time (year-scaled) radio signal might originate from the ejecta-medium interaction as the ejecta enters the Sedov-Taylor phase \citep{nak11}.

Although BNS mergers have been proposed as one of the possible progenitors of short GRBs over the past three decades \citep{pac86, eich89, nar92, tut92, moch93, bog07} and there is a plenty of indirect evidence for such a scenario \citep[e.g., for reviews see][]{nak07, ber14}, a conclusive proof remains lacking. It is generally believed that the detection of GW emission can provide a unique way to verify this scenario. However, the advanced LIGO/Virgo GW detection horizon of BNS mergers is only about one hundred mega-parsecs \citep{aba10, mar16} and short GRBs rarely fall into this close distance range.

Luckily, the first strong evidence of GW170817 associated with GRB170817A was discovered very recently \citep{abb17c}, benefited from its relatively close distance. It is beyond doubt a landmark in multi-messenger astronomy and can greatly enhance our understanding of BNS mergers. The host galaxy associated with GW170817/GRB170817A is found to be NGC 4993 with a luminosity distance of $D_L\simeq40\,\rm Mpc$ \citep{abb17d, hjo17}. Observationally, this GRB has a duration of $T_{90}\sim2\,\rm s$, an isotropic-equivalent $\gamma$-ray energy of $E_{\rm iso}\sim 4.6\times 10^{46}\,{\rm erg}$, and an isotropic-equivalent peak luminosity of $L_{\rm iso,peak}\sim\rm 1.7\times10^{47}\,\rm erg\,s^{-1}$ \citep{fermi17, zhangBB17}, which shows that this GRB is a few orders of magnitude less energetic than a typical (high-luminosity) short GRB \citep{kie17}.

According to the statistic analysis of the luminosity function and burst rate of short GRBs \citep{sun15, ghir16}, nearby low-luminosity short GRBs (with luminosity, e.g., $L_{\rm iso}<10^{48}\,\rm erg\,s^{-1}$) may be much more numerous than normal ones and we have a greater chance to detect them. Generally, low-luminosity short GRBs could originate from less powerful central engines. Nevertheless, there is another possibility that we are off-axis and only observe the side emission of a normal short GRB since its detection probability should be higher than that of on-axis emission \citep{laz17b}. For instance, the side emission from an off-axis short GRB with a structured jet has been discussed as possible EM counterparts to GWs \citep{kath17} and also several other radiation components such as the cocoon emission have been proposed as possible counterparts in previous works \citep{got17, lamb17, laz17a, laz17b, jin17}. This kind of side emission should be much fainter than the on-axis jet emission from an observational point of view \citep[e.g.][]{yam02, yam03}. The fact that GRB170817A has a typical peak energy $E_p$ \citep{fermi17} would not conflict with these models, since the prompt emission mechanism is unknown and the observed gamma-rays could either arise from the emission of the jet scattered to a wide angle \citep{kis17} or just from the emission produced as the cocoon breaks out of the ejecta \citep{got17b, kas17}.

Based on the above argument, we here consider several cases in which the viewing angle $\theta_{\rm v}$ varies. We carry out calculations of multi-wavelength afterglow emission with different viewing angles under the assumption of a universally-structured jet and a two-component jet respectively, and then make a comparison with that of an intrinsically less-energetic quasi-isotropic fireball. Our results show that such three types of model are distinguishable and can be tested by multi-wavelength follow-up observations. We apply these models to GRB170817A and find that the two-component jet model with reasonable parameters matches the observations better than the structured jet model does. Furthermore, we explore the kilonova emission from the BNS merger ejecta and constrain the ejecta parameters with the observations.

This paper is organized as follows. In Section 2 we introduce the universally-structured jet model and the two-component jet model, and calculate the off-axis afterglow emission. Then, we present the method of calculations for the kilonova emission in Section 3. Section 4 shows our results for the two jet models and gives a comparison with an intrinsically less-energetic quasi-isotropic fireball. Section 5 is an application to the very recently-discovered GW170817/GRB170817A. Lastly, we draw conclusions and provide a summary in Section 6.

\section{Off-axis afterglows}
In this section, we consider a structured jet with a lateral distribution of kinetic energy per solid angle $\varepsilon(\theta)$. This kind of jet may form during the propagation of the jet inside the ejecta, which gives rise to shocks at the jet head \citep{nag14, nak17}. The relativistic shocked jet material forms an inner cocoon, which is wrapped by an outer cocoon composed of mildly-relativistic shocked ejecta \citep{got17, nak17, laz17a, laz17b}. Although there is some mixing between them, the cocoon is far from isotropy \citep{nak17, laz17b}. Thus, the overall uniform jet core plus structured cocoon system can be named as a structured jet, of which the kinetic energy per solid angle and the initial Lorentz factor are assumed to be \citep{dai01, zhang02, ros02, kum03}
\beq
\varepsilon(\theta)\equiv\frac{dE}{d\Omega}=\begin{cases}
\varepsilon_0, & \text{if} \,\,\theta\leq\theta_c, \\
\varepsilon_0(\theta/\theta_c)^{-k}, & \text{if}\,\, \theta_c<\theta<\theta_m,
\end{cases}
\label{eq:Edis}
\enq
\beq
\Gamma_0(\theta)=\begin{cases}
\Gamma_0, & \text{if} \,\,\theta\leq\theta_c, \\
\Gamma_0(\theta/\theta_c)^{-s}, & \text{if}\,\, \theta_c<\theta<\theta_m,
\end{cases}
\label{eq:Gamdis}
\enq
where the typical half opening angle of short GRBs $\theta_c\simeq 0.1$ \citep[which is marginally consistent with the median opening angle given by][]{fong15} and the maximum angle $\theta_m=4\theta_c$ are assumed. The index $k$ can be deduced from the luminosity distribution of local event rate density $\rho_0(>L)$. On the one hand, the local event rate density of short GRBs can be fitted by a power-law $\rho_0(>L)\propto L^{-\lambda}$ with $\lambda\sim0.7$ \citep{sun15}. Since $\rho_0(>L)\propto\Omega(>E)\simeq\pi\theta^2$ for similar durations of prompt emission, we can get $L\propto\theta^{-2/\lambda}$. On the other hand, the isotropic-equivalent luminosity $L\propto 4\pi\times dE/d\Omega\propto \theta^{-k}$. Therefore, $k=2/\lambda\simeq2.86$. In this paper, we adopt $k=3$ as a nominal value. Generally, the relationship of indexes $s$ and $k$ can be deduced from some empirical relations. Observationally, the relationship between the initial Lorentz factor and isotropic-equivalent energy is approximated by $\Gamma_0\propto E_{\gamma,\rm iso}^{1/4}$ \citep[e.g.][]{zhang02b, liang10, lv12}. Thus, since $L\propto E_{\gamma,\rm iso}$, we have $s\simeq k/4$ in this work.

Similarly, a two-component jet can be described by the following angular distributions \citep[e.g.,][]{vla03, huang04,lamb17}
\beq
\varepsilon(\theta)=\begin{cases}
\varepsilon_{\rm in}, & \text{if} \,\,\theta\leq\theta_c, \\
\varepsilon_{\rm out}, & \text{if}\,\, \theta_c<\theta<\theta_m,
\end{cases}
\label{eq:Edis2}
\enq
\beq
\Gamma_0(\theta)=\begin{cases}
\Gamma_{\rm in}, & \text{if} \,\,\theta\leq\theta_c, \\
\Gamma_{\rm out}, & \text{if}\,\, \theta_c<\theta<\theta_m,
\end{cases}
\label{eq:Gamdis2}
\enq
where $\varepsilon_{\rm in}$ and $\varepsilon_{\rm out}$ and $\Gamma_{\rm in}$ and $\Gamma_{\rm out}$ represent the kinetic energies and Lorentz factors of the inner fast spine and outer slow sheath respectively.

For an off-axis viewing angle $\theta_{\rm v}$, the infinitesimal patch of the emission region at $(r,\theta,\phi)$ makes an angle $\alpha$ with respect to the observer, which is given by \citep{kath17}
\beq
\cos\alpha=\cos\theta_{\rm v}\cos\theta+\sin\theta_{\rm v}\sin\theta\cos\phi.
\label{eq:ang}
\enq
Assuming that the jet expands outward in a homogeneous medium with a typical number density $n\sim 10^{-2}\,\rm cm^{-3}$ for short GRBs \citep{fong15}, the evolution of the bulk Lorentz factor $\Gamma$ can be obtained in the same way as previous works \citep[e.g.][]{bm76, huang99, dai01}. In this paper, we adopt the generic dynamics of a jet following \cite{huang00} without considering any lateral expansion of the jet. The radius and the time $t^{\prime}$ in the jet's comoving frame can be expressed by
\bea
\frac{dR}{dt}=\frac{c\beta}{1-\beta\cos\alpha},
\ena
and
\bea
\frac{dt^{\prime}}{dt}=\frac{1}{\Gamma(1-\beta\cos\alpha)},
\label{eq:Rtcom}
\ena
where $\beta\equiv(1-1/\Gamma^2)^{1/2}$ and $t$ is the observed time.

Now we can calculate synchrotron radiation of the electrons accelerated by a forward shock produced due to an interaction of the jet with its ambient medium. Assuming the electrons have a power-law distribution $dn_e/d\gamma_e\propto \gamma_e^{-p}$, the minimum electron Lorentz factor is then $\gamma_m=[(p-2)/(p-1)]\epsilon_e(m_p/m_e)\Gamma$, where $\epsilon_e$ is a fraction of the post-shock energy density converted to electrons and the spectral index of the electron energy distribution $p=2.5$ is adopted as a nominal value. The cooling Lorentz factor is $\gamma_c=6\pi m_ec/(\sigma_TB^{\prime2}t^{\prime})$, where the magnetic field strength in the shocked medium is given by $B^{\prime}=[32\pi\epsilon_B\Gamma(\Gamma-1)nm_pc^2]^{1/2}$ with $\epsilon_B$ being a fraction of the post-shock energy density converted to a magnetic field. In this paper, we adopt typical equipartition factors $\epsilon_e=0.1$ and $\epsilon_B=0.01$ for short GRBs \citep{fong15}. With these parameters, we can calculate the typical frequency $\nu_m^{\prime}$ and the cooling frequency $\nu_c^{\prime}$. According to the relative values of the two frequencies, the spectrum without synchrotron self absorption (SSA) can be written \citep{sari98}. The SSA frequency $\nu_a^{\prime}$ can be obtained by equaling the blackbody luminosity at the Rayleigh-Jeans end with the synchrotron luminosity. At last, we can write down the complete differential luminosity $dL_{\nu'}^{\prime}/d\Omega'$ in the jet's comoving frame \citep[e.g.][]{dai01, xiao17}.

The observed total flux density of the off-axis afterglow is then given by \citep{dai01, gra02, kath17}
\beq
F_\nu=\int_0^{\theta_m}d\theta\int_0^{2\pi}d\phi\frac{dL_{\nu^{\prime}}^{\prime}/d\Omega'}{4\pi D_L^2\Gamma^3(1-\beta\cos\alpha)^3},
\enq
where $D_L$ is the luminosity distance of the source to an observer. Note that we should integrate on the equal arrival time surface that is determined by $t=\int(1-\beta\cos\alpha)/(c\beta)dR\equiv\rm constant$ \citep{wax97, pan98, sa98, huang00, mod00}.

\section{Kilonovae}
The neutron-rich ejecta produced during a BNS merger
undergoes rapid neutron capture ($r$-process) nucleosynthesis. The
radioactive decay of these heavy nuclei is able to power a day-to-week-long
kilonova \citep{li98,kul05,met10,kas13,tan13,met17}.

%The fundamental equation of the kilonova evolution is the first law of thermodynamics, which can be written as%
%\begin{equation}
%\frac{dE_{\text{int}}}{dt}=-P\frac{dV_{\rm ej}}{dt}+\dot{Q}-L_{\text{MN}},
%\end{equation}
%where $E_{\text{int}}$ is the internal energy of the ejecta, $P$ is
%the pressure, $V_{\rm ej}$ is the volume of the ejecta, and $\dot{Q}$ is the heating rate. Radiation energy
%dominates over gas energy. In this case, the pressure is given by
%\begin{equation}
%P=\frac{1}{3}\frac{E_{\text{int}}}{V_{\rm ej}}.
%\end{equation}

The density distribution of the ejecta can be obtained from numerical
simulations. The geometry structure of the ejecta can be modeled as a partial sphere in the
latitudinal and longitudinal direction \citep{kyu13,kyu15}.  We
assume a homologous expansion inside the ejecta, so the
density of the ejecta is \citep{kaw16}
\begin{equation}
\rho \left( v,t\right) =\frac{M_{\text{ej}}}{2\phi _{\text{ej}}\theta _{%
\text{ej}}\left( v_{\max }-v_{\min }\right) }v^{-2}t^{-3},
\end{equation}
where $v_{\min }$ and $v_{\text{max}}$ are the minimum and
maximum velocities of the ejecta respectively, $\theta _{\text{ej}}$
is the polar opening angle, and $\phi _{\text{ej}}$ is the azimuthal
opening angle. Here, we adopt $v_{\min }=0.02c,$ and $%
v_{\text{max}}=2v_{\text{ej}}-v_{\min }$. For a BNS merger, there exists a
linear correlation between $\theta _{\text{ej}}$ and $\phi
_{\text{ej}}$ \citep{die17},
\begin{equation}
\phi _{\text{ej}}=4\theta _{\text{ej}}+\frac{\pi }{2}.
\end{equation}

We assume that the kilonova is powered radioactively, without an additional energetic engine such as a stable strongly-magnetized millisecond pulsar as suggested in the literature \citep[e.g.,][]{dai2006,yu2013}. The heating rate of $r$-process ejecta can be approximated by a power law \citep{kor12,tana16}
\begin{equation}
\dot{Q}\approx M_{\text{ej}}\epsilon _{0} \left( \frac{t }{\text{day}}\right)^{-\alpha},
\end{equation}%
where we adopt $\epsilon _{0}=1.58\times 10^{10}\,\rm erg\, s^{-1}\,{\rm g}^{-1}$ and $\alpha=1.3$ following \cite{die17}.

The bolometric luminosity of kilonova is approximated by
\citep{kaw16,die17}
\begin{equation}
L_{\text{MN}}=\left( 1+\theta _{\text{ej}}\right) \dot{Q}\epsilon _{\text{th}}\times
\left\{
\begin{array}{ll}
t/t_{c}, & {\rm if}\,\,t\leq t_{c}, \\
1, & {\rm if}\,\,t>t_{c},%
\end{array}%
\right.
\end{equation}
where the factor $\left( 1+\theta _{\text{ej}}\right) $ indicates the
contribution from an effective radial edge. $\epsilon _{\text{th}}$ is the
thermalization efficiency introduced in \cite{met10}, and we adopt $\epsilon _{\text{th}}=0.5$ as a nominal value.  The critical time $t_{c}$
at which the expanding ejecta becomes optically thin \citep{kaw16} is
\begin{equation}
t_{c}=\left[ \frac{\theta _{\text{ej}}\kappa M_{\text{ej}}}{2\phi _{\text{ej}%
}\left( v_{\max }-v_{\min }\right) c}\right] ^{1/2}.
\end{equation}
For $t<t_{c}$, the mass of the photon-escaping region is
$M_{\text{obs}}(t)=M_{\text{ej}}(t/t_c)$. At $t=t_c$, the whole
region of the ejecta becomes transparent. \cite{kas13} and
\cite{bar13} found that the opacity of $r$-process ejecta,
particularly the lanthanides, is much higher than that for Fe-peak
elements, with $\kappa \sim 10-100$ cm$^{2}$ g$^{-1}.$
\cite{kaw16} and \cite{die17} found that the bolometric light curve of a kilonova
in the analytic model mentioned above can well match the results of
radiation-transfer simulations performed in \cite{tana13}.

Assuming that the spectrum of the kilonova emission is approximated by a blackbody, the effective
temperature can be written as
\begin{equation}
T_{\text{eff}}=\left( \frac{L_{\text{MN}}}{\sigma
_{\text{SB}}S}\right) ^{1/4},
\end{equation}
where $\sigma_{\rm SB}$ is the Stephan-Boltzmann constant and $S=R_{\rm ej}^2\phi_{\rm ej}$
is the emitting area with $R_{\rm ej}\simeq v_{\text{max}}t$ being the radius of the latitudinal edge. The observed flux at
photon frequency $\nu $ can be calculated by
\begin{equation}
F_{\nu ,\text{MN}}=\frac{2\pi h\nu ^{3}}{c^{2}}\frac{1}{\exp \left( h\nu /k_{%
\text{B}}T_{\text{eff}}\right) -1}\frac{R_{\rm ej}^{2}}{D_{L}^{2}},
\end{equation}
where $h$ is the Planck constant and $k_{\rm B}$ is the Boltzmann
constant.

\section{Theoretical Results}
Figure 1 shows our theoretical light curves of the structured jet model for different viewing angles. We consider a typical short GRB with jet core energy $\varepsilon_0=10^{50}\,\rm erg/sterad$ and Lorentz factor $\Gamma_0=300$, located at a close distance $D_L=40\,\rm Mpc$. With the increase of the viewing angle, the peak luminosity decays and the X-ray light curve shifts to earlier times until $\theta_{\rm v}$ becomes larger than $\theta_m$, which is different from previous works that assume $s=k$ \citep[e.g.][]{mod00,huang00,gra02,lamb17}. The reason for this result is that the light curves peak when the break frequencies ($\nu_m^{\prime}$ and $\nu_c^{\prime}$) cross the observed frequency \citep{sari98}. Since $\nu_m^{\prime}=\gamma_m^2eB^{\prime}/(2\pi m_e c)=1.95\times10^7n_{-2}^{1/2}\Gamma(\Gamma-1)^{5/2}\,\rm Hz$ and $\nu_c^{\prime}=\gamma_c^2eB^{\prime}/(2\pi m_e c)=2.85\times10^{31}n_{-2}^{-3/2}\Gamma^{-3/2}(\Gamma-1)^{-3/2}t^{\prime-2}\,\rm Hz$ for the  parameters taken in Section 2, we find that $\nu_m^{\prime}<\nu_c^{\prime}$ is always satisfied so the synchrotron emission is in the slow cooling regime. After converting the observed frequencies into the comoving frame, we see that
$\nu_m^{\prime} < \nu_X^{\prime} < \nu_c^{\prime}$ is always established, while initially $\nu_{\rm r-band}^{\prime} < \nu_m^{\prime} < \nu_c^{\prime}$ but soon turns into $\nu_m^{\prime} < \nu_{\rm r-band}^{\prime} < \nu_c^{\prime}$, and initially $\nu_{\rm radio}^{\prime} < \nu_m^{\prime} < \nu_c^{\prime}$ but turns into $\nu_m^{\prime} < \nu_{\rm radio}^{\prime} < \nu_c^{\prime}$ at a much later time. This gives rise to different peak times of different bands. The light curves of r-band are shown in Figure 1(b). Solid lines are corresponding to afterglow emission, and dashed and dotted lines to kilonova emission. The theoretical flux of the kilonova signal depends on the kinetic energy and velocity of the ejecta. Numerical simulations have suggested that the ejecta has typical mass $10^{-4}-10^{-2}\,M_{\odot}$ and velocity $0.1-0.3c$ \citep[e.g.][]{nag14}. Thus we consider two masses $10^{-3}M_{\odot}$ (magenta) and $10^{-2}M_{\odot}$ (red), and velocities $0.1c$ (dotted) and $0.3c$ (dashed), so we have four combinations. For large viewing angles, the kilonova signal probably dominates over the afterglow. Therefore, if the kilonova component can be extracted in optical-infrared follow-up observations, it will help constrain parameters such as the viewing angle and the ejecta mass and velocity. For completeness, we plot the light curves of the radio band ($\nu=5\,\rm GHz$) in Figure 1(c). Since the wide-angle structured jet (including its cocoon) sweeps up its ambient medium at early times, there might be no medium leftover and thus the ejecta will possibly expand freely with a nearly constant velocity. Thus, we neglect any emission from an interaction of the ejecta with its ambient gas in a year-scale period after the merger \citep{nak11}. The time evolution of the afterglow spectrum is shown in Figure 1(d) for the $\theta_{\rm v}=4\theta_c$ case.

The theoretical results in the two-component jet model shown in Figure 2 are very different from those of the structured jet model. The line styles in this figure are the same as those in Figure 1. For an off-axis observer, the afterglow emission is dominated by the wide component at early times. The relevant parameters are $\Gamma_{\rm in}=300,\,\Gamma_{\rm out}=30, \varepsilon_{\rm in}=10^{50}\,\rm erg/sterad$ and $\varepsilon_{\rm out}=10^{48}\,\rm erg/sterad$. The emission from the narrow component generally shows up at times later than $10^5\,\rm s$. The ratio of peak luminosities between the wide and narrow component depends on the ratio of their energy and viewing angle. With the increase of $\theta_{\rm v}$, the peak time delays and the peak luminosity decays.

However, there is still a possibility that an observed low-luminosity burst is not due to a large viewing angle, and instead it arises from an intrinsically less-energetic quasi-isotropic fireball. We need to consider its afterglow emission for completeness. The structured jet model can be easily generalized to an isotropic fireball case if we set index $k=0$ and opening angle $\theta_m=\pi$ in Equation ({\ref{eq:Edis}). Since the kinetic energy per solid angle along the line of sight in the structured jet model can be estimated by $\varepsilon_0/\varepsilon_{\rm obs}=(\theta_c/\theta_{\rm v})^{-k}$, to make a direct comparison with one of the previous cases (e.g., $\varepsilon_0=10^{50}\,\rm erg/sterad,\,\theta_{\rm v}=4\theta_c$), we assume a fireball with isotropic kinetic energy $E_{\rm iso}\sim4\pi\times10^{50}\times4^{-3}\,{\rm erg}\sim2.0\times10^{49}\,{\rm erg}$. The corresponding X-ray, r-band, radio light curves and spectral evolution are shown in Figures 3. Different lines represent different medium densities, ranging from $n=10^{-4}-1\,\rm cm^{-3}$. As is expected, the flux level drops with the decrease of $n$. Note that the radio light curve shape varies with medium density because there is a crossing between $\nu_a^{\prime}$ and $\nu_{\rm radio}^{\prime}$ for higher densities (in the cases of $n=1$ and $0.1\,\rm cm^{-3}$) while $\nu_a^{\prime}<\nu_{\rm radio}^{\prime}$ always holds for densities lower than $10^{-2}\,\rm cm^{-3}$. We can clearly see that the observed afterglow emission of an intrinsically less-energetic fireball is very different from that of an intrinsically powerful off-axis short GRB discussed above. In particular, comparing the yellow solid line in Figure 1(a) with the blue solid line in Figure 3(a), we can see that the peak time and peak luminosity differ (about two order of magnitude) for these two types of model. Similar differences can be found in r-band and radio band. Also, the quasi-isotropic kilonova signal may be different since intrinsically-fainter short GRBs are likely accompanied by less-energetic ejecta, so the kilonova should be dimmer. The spectral evolution with time is also different from each other in the two types of model if we compare Figure 3(d) with Figure 1(d). All of these results would be testable by multi-wavelength follow-up observations.

\section{Application to GW170817/GRB170817A}

In this section, we try to fit the multi-wavelength observational data \citep[for a complete collection, see][]{abb17d} with the above models and the results are shown in Figure 4 and Figure 5. The relevant fitting parameters are given in Table 1 and Table 2. The X-ray upper limits are given by {\em Swift-XRT} and {\em NuSTAR} \citep{swinupaper17}, while the two  detections of {\em Chandra} are indicated by the red datapoints \citep{chan17}. For the optical band, we choose r-band to fit and the data are collected in the literature \citep{and17, arc17, cou17, dro17, kil17, pian17, sha17, sma17}. The six radio datapoints ($\nu=3\,\rm GHz$) are taken from \cite{hall17} and could give tight constraints on models. The quasi-isotropic fireball model is ruled out because the early X-ray flux is over-estimated if the X-ray light curve is required to pass through the {\em Chandra} datapoint. The structured jet model can account for the {\em Chandra} X-ray data without violating the {\em Swift-XRT} and {\em NuSTAR} upper limits only if $\theta_{\rm v}>\theta_m$. Obviously, the two-component model works well for X-ray emission. For the optical band, we use a Markov Chain Monte Carlo (MCMC) approach and the r-band data can be well fitted by the kilonova component, while the afterglow emission is always sub-dominant, which is shown in Figure 4(b) and Figure 5(b). The MCMC best-fitting parameters obtained from the corner plot shown in Figure 6 for the kilonova component are $\tilde{M}_{\text{ej}}=0.026\pm 0.0016$ ($2\sigma$) and $v_{\rm ej}=(0.12 \pm 0.015)c$ ($2\sigma$), where $\tilde{M}_{\text{ej}}=(\kappa/10\,{\rm{cm}^{2} \,{g}}^{-1})\times (M_{\rm ej}/M_{\odot})$ is defined. The radio data can provide the tightest constraint on the models. The comparisons of the fitting results with radio observations are given in Figure 4(c) and Figure 5(c). Generally the two-component jet model gives the better-fitting quality than the structured jet model. However, the radio data are still dimmer than model predictions in all the cases, indicating that there might be an extra component leading to the delayed X-ray emission (e.g. the contribution of a reverse shock). Note that there are some degeneracies in parameters (energy, medium density, viewing angle, etc) and better-fitting quality can be achieved through fine tuning.

\section{Conclusions}
\label{sec:disc}
The discovery of multi-wavelength EM signals associated with GW170817 marks the beginning of a new era in multi-messenger time-domain astronomy. In this work we have first re-investigated both an afterglow and a kilonova which are associated with a nearby low-luminosity short GRB from a BNS merger, under the assumption of a universally-structured jet and a two-component jet. We then tried to apply the models to GW170817/GRB170817A. In general, the isotropic fireball model is ruled out because it is fully inconsistent with the early X-ray upper limits and radio data. The structured jet model may explain the late-time X-ray emission but predict a radio flux much higher than observed. The multi-wavelength observational data could be well fitted by the two-component jet model and all the relevant parameters are within their reasonable ranges, although further fine tuning is needed.

Generally, detecting a low-luminosity short GRB is estimated to be much easier in our local universe than a normal one because the former has a much greater occurrence rate than the latter does. With the upgrade of the advanced LIGO/Virgo detectors and the improvement of the all sky transient survey, this kind of association will become more and more common in the future. We have considered two possibilities that either a faint short GRB (like GRB170817A) is intrinsically low-luminosity and quasi-isotropic or it is just due to off-axis jet emission. We have shown that the properties of afterglow emission in these cases are obviously different. The light curves rise slower and peak at a later time for the off-axis case \citep[e.g.][]{murg17}. The spectrum is also different at any given time. Furthermore, if we assume the kinetic energy of the ejecta is proportional to that of the jet, the kilonova signal in the less-energetic fireball case should be much fainter than that of the off-axis powerful short GRB case. With multi-wavelength follow-up observations of a local low-luminosity short GRB, we can distinguish between these models very soon if such a kind of association is confirmed again in the future. In addition, several key parameters can be constrained such as the viewing angle, the ejecta mass, the ejecta velocity, and the ambient medium density, all of which would help reveal the mystery of short GRBs. For GW170817/GRB170817A, the ejecta parameters obtained in this paper are
$(\kappa/10\, {\rm{cm}^{2} \,{g}}^{-1})\times({M_{\rm ej}}/M_{\odot})=0.026\pm 0.0016$ ($2\sigma$) and $v_{\rm ej}=(0.12 \pm 0.015)c$ ($2\sigma$) by considering the kilonova component. These parameters are well consistent with numerical simulations of BNS mergers.

\acknowledgements
We thank the referee and Bing Zhang for helpful comments and constructive suggestions. This work was supported by the National Basic Research Program of China (973 Program grant 2014CB845800), the Strategic Priority Research Program ``Multi-waveband gravitational wave Universe'' (grant No. XDB23040000) of the Chinese Academy of Sciences, and the National Natural Science Foundation of China (grant No. 11573014, 11673068, 11433009, and 11725314).  XFW was also partially supported by the Youth Innovation Promotion Association (No. 2011231) and the Key Research Program of Frontier Sciences (QYZDB-SSW-SYS005).

%\clearpage

\clearpage
\begin{figure}
\centering
\subfigure[]{
\begin{minipage}[b]{0.45\textwidth}
\includegraphics[width=1\textwidth]{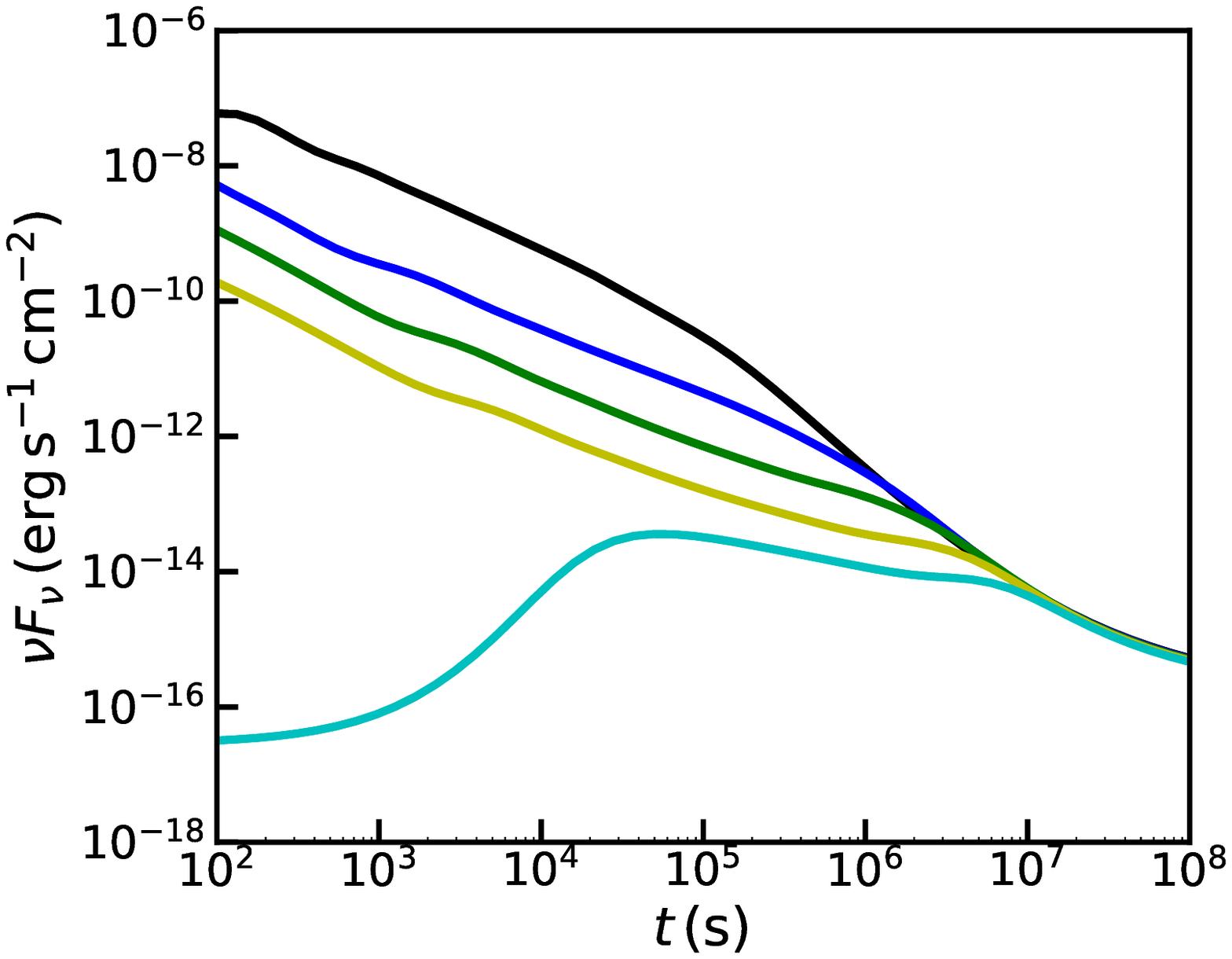}
\label{fig1:subfig:a}
\end{minipage}
}
\subfigure[]{
\begin{minipage}[b]{0.45\textwidth}
\includegraphics[width=1\textwidth]{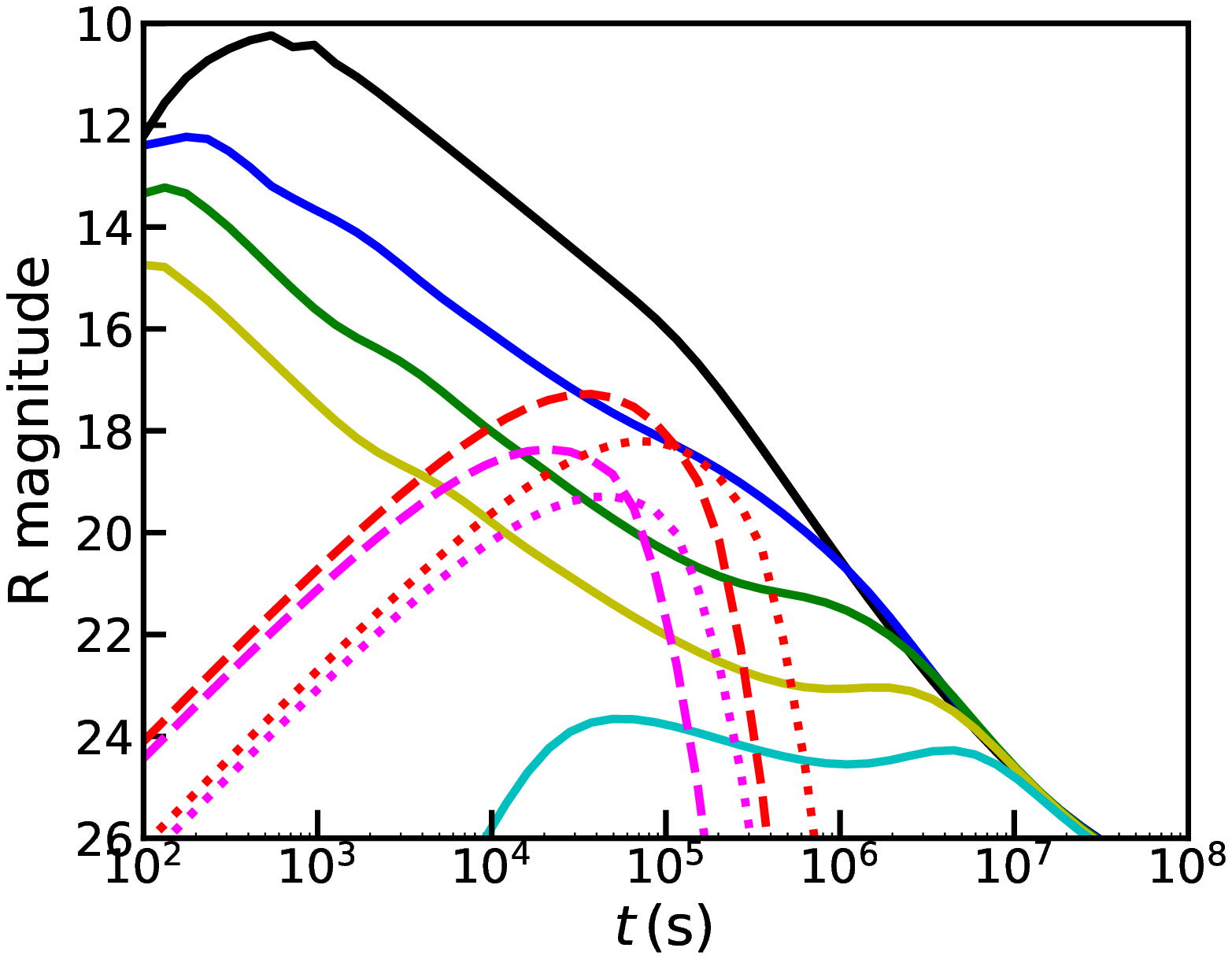}
\label{fig1:subfig:b}
\end{minipage}
}
\subfigure[]{
\begin{minipage}[b]{0.45\textwidth}
\includegraphics[width=1\textwidth]{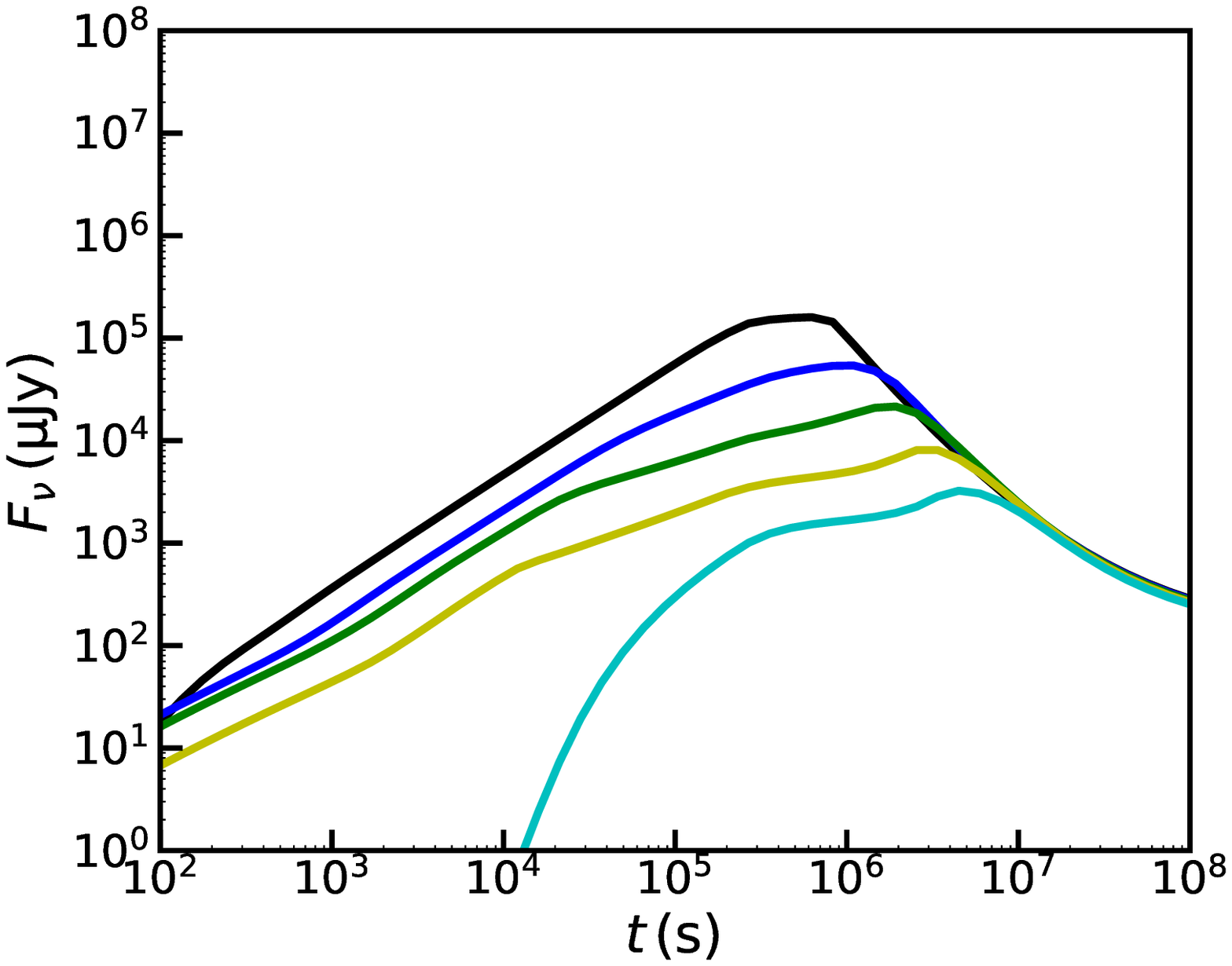}
\label{fig1:subfig:c}
\end{minipage}
}
\subfigure[]{
\begin{minipage}[b]{0.45\textwidth}
\includegraphics[width=1\textwidth]{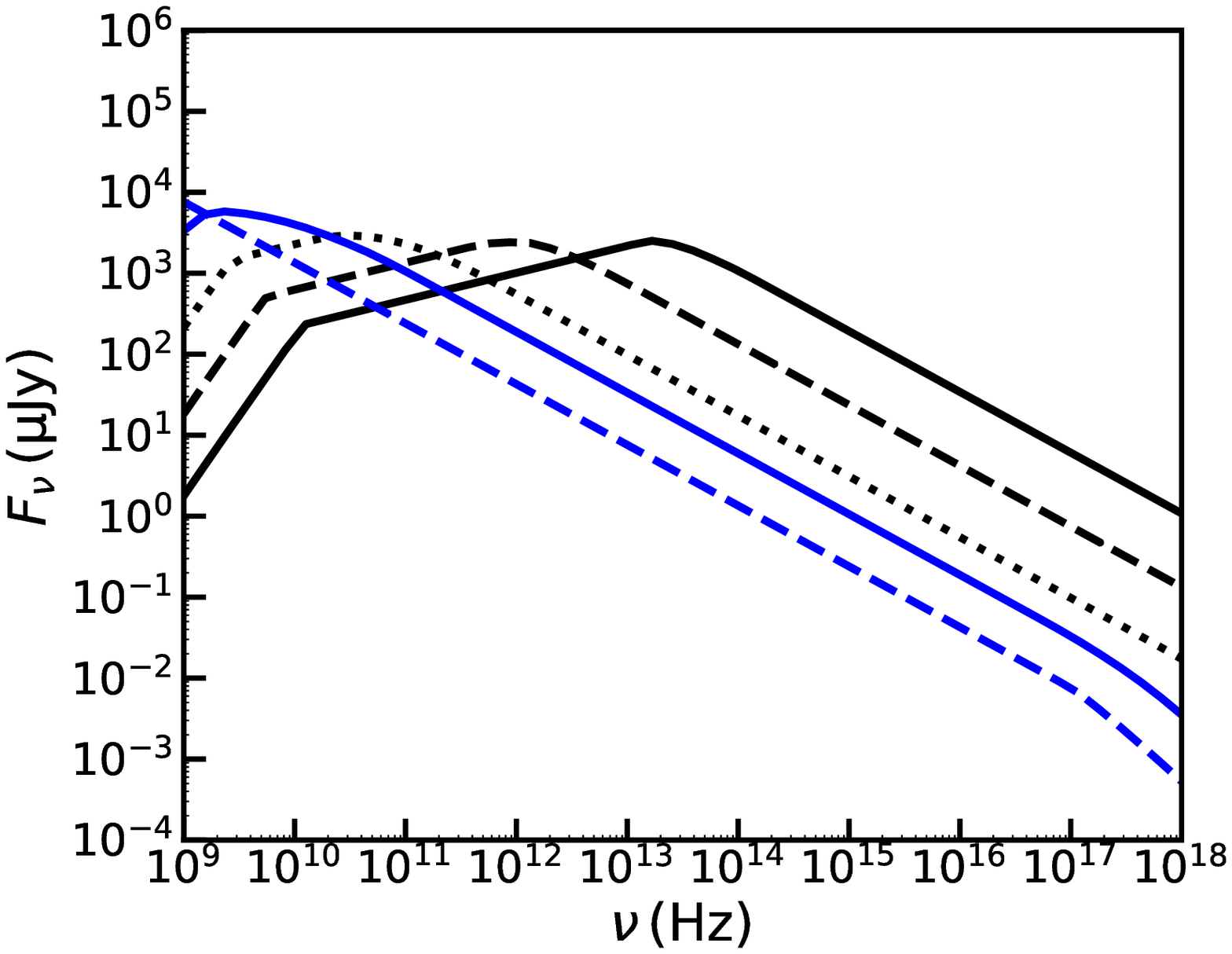}
\label{fig1:subfig:d}
\end{minipage}
}
\caption{The theoretical results in the structured jet model. Panel (a): The X-ray light curves for different viewing angles. The black, blue, green, yellow and cyan solid lines are corresponding to $\theta_{\rm v}=0,\, 2\theta_c,\, 3\theta_c,\,4\theta_c,$ and $ 5\theta_c$ respectively. The medium density is taken as $n=10^{-2}\,\rm cm^{-3}$. Panel (b): The r-band magnitude for different viewing angles. The black, blue, green, yellow and cyan solid lines are corresponding to the afterglow emission of $\theta_{\rm v}=0,\, 2\theta_c,\, 3\theta_c,\,4\theta_c,$ and $5\theta_c$ respectively. The four kilonova signals for $\theta_{\rm ej}=\pi/4$ can be distinguished by colors (magenta for $M_{\rm ej}=10^{-3}M_{\odot}$ and red for $M_{\rm ej}=10^{-2}M_{\odot}$) and line styles (dotted for $v_{\rm ej}=0.1c$ and dashed for $v_{\rm ej}=0.3c$. Panel (c): The radio ($\nu=5\,\rm GHz$) light curves for different viewing angles. The line styles are the same as in panel (a). Panel (d): The spectrum evolution for the viewing angle $\theta_{\rm v}=4\theta_c$ case. The black solid, dashed, dotted, blue solid, and blue dashed lines represent the spectra at $t=10^3,\,10^4,\,10^5,\,10^6,$ and $10^7\,\rm s$ respectively.
\label{Figure 1}}
\end{figure}

\begin{figure}
\centering
\subfigure[]{
\begin{minipage}[b]{0.45\textwidth}
\includegraphics[width=1\textwidth]{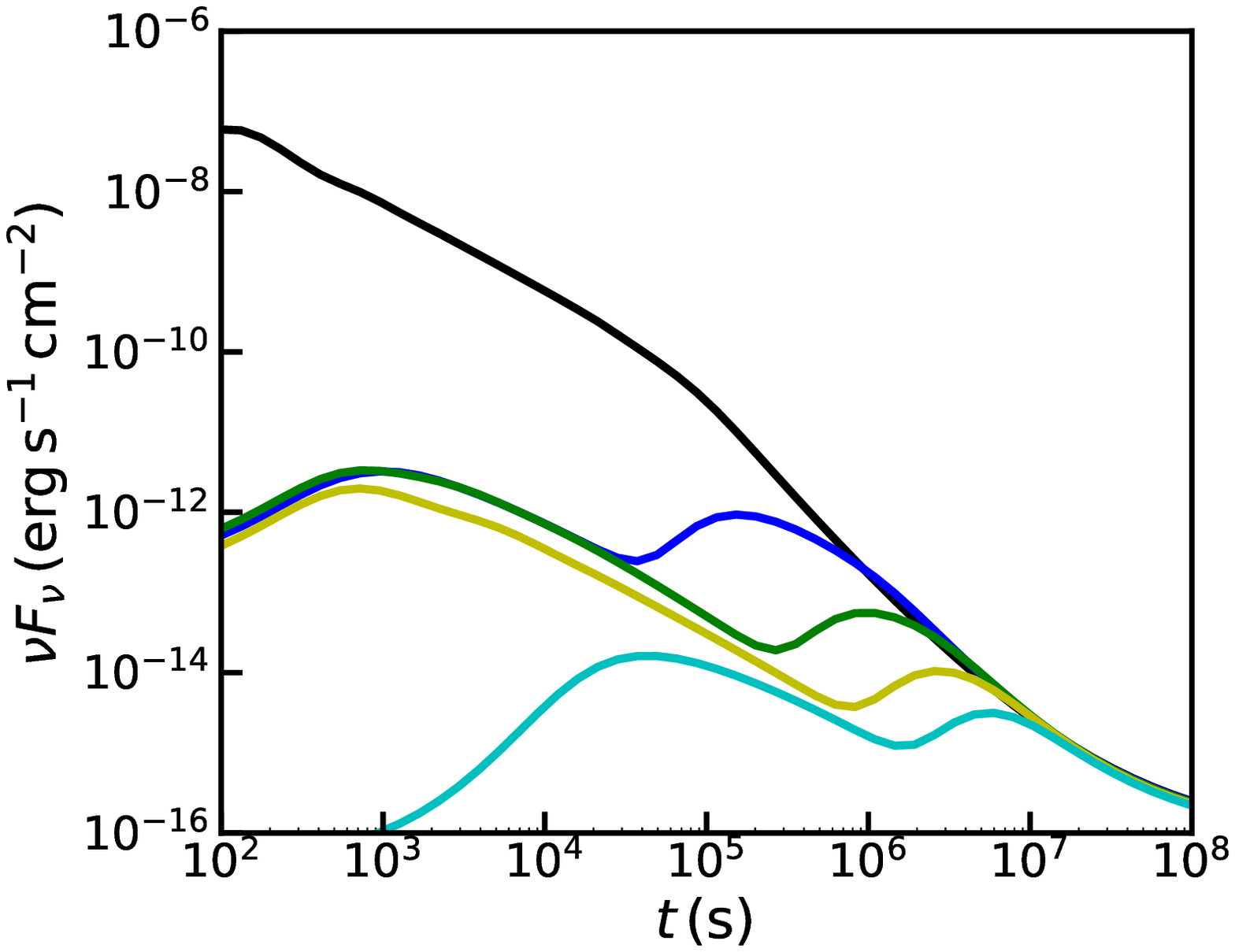}
\label{fig2:subfig:a}
\end{minipage}
}
\subfigure[]{
\begin{minipage}[b]{0.45\textwidth}
\includegraphics[width=1\textwidth]{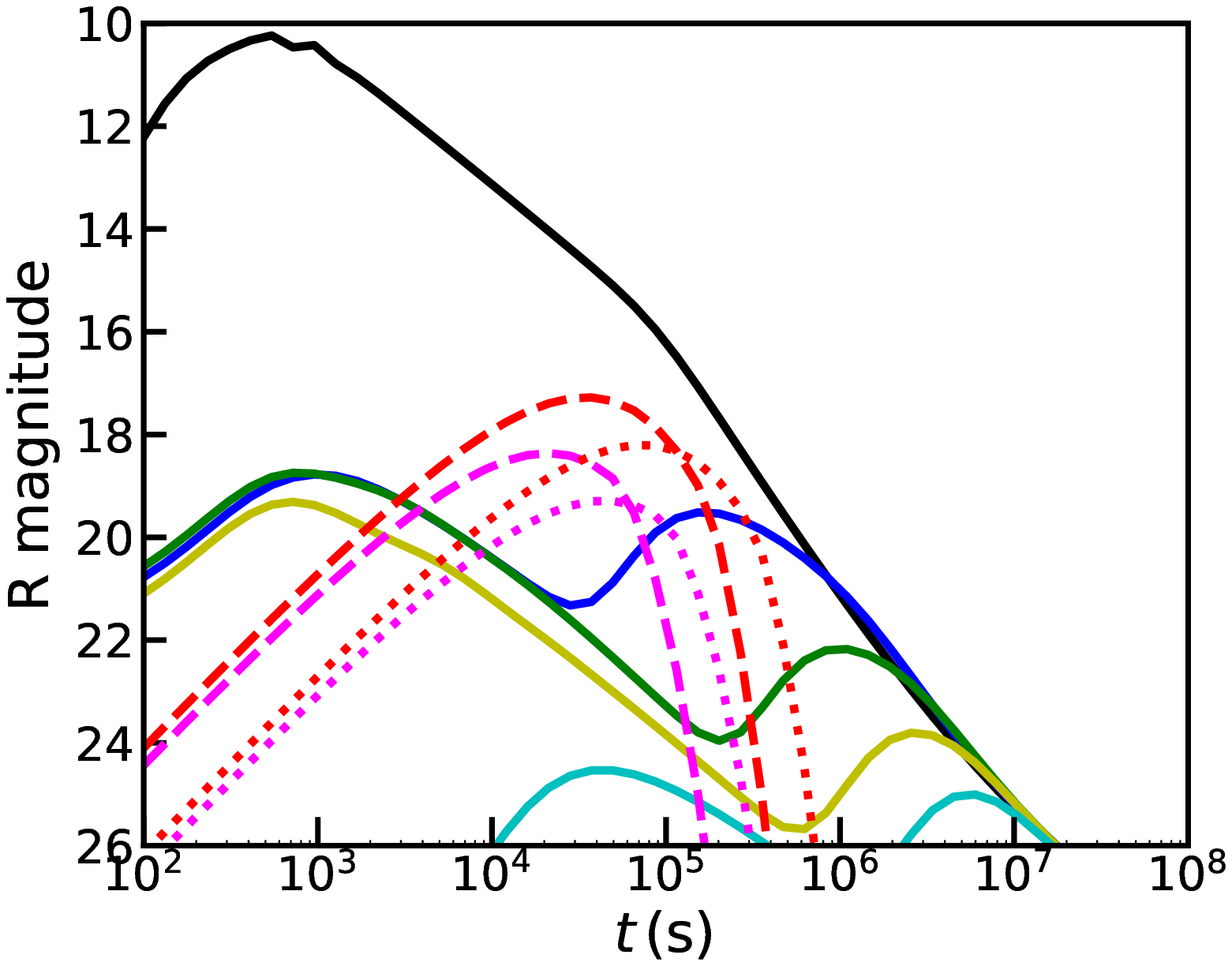}
\label{fig2:subfig:b}
\end{minipage}
}
\subfigure[]{
\begin{minipage}[b]{0.45\textwidth}
\includegraphics[width=1\textwidth]{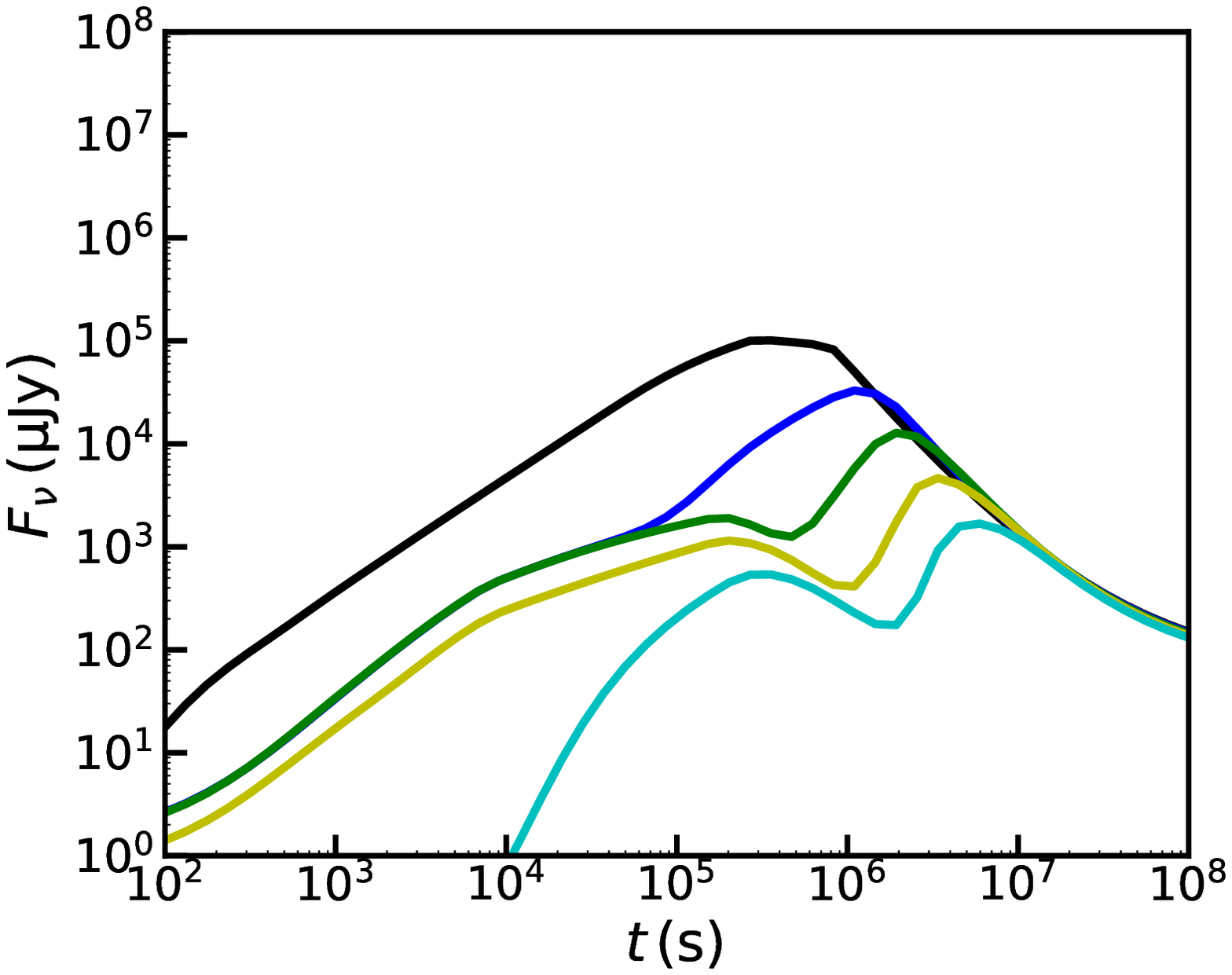}
\label{fig2:subfig:c}
\end{minipage}
}
\subfigure[]{
\begin{minipage}[b]{0.45\textwidth}
\includegraphics[width=1\textwidth]{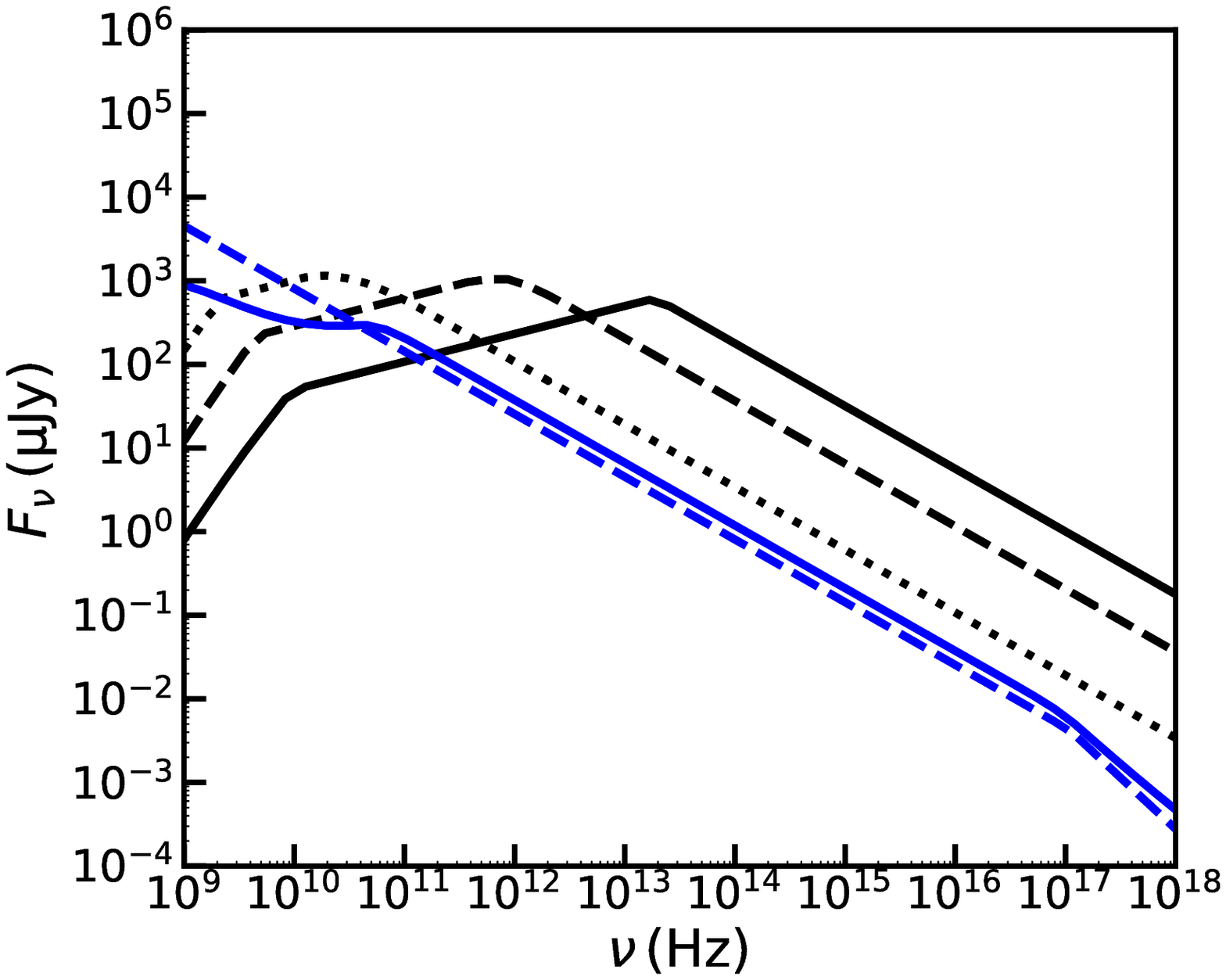}
\label{fig2:subfig:d}
\end{minipage}
}
\caption{The theoretical results in the two-component jet model. Panel (a): The X-ray light curves for different viewing angles. The line styles are the same as in Figure 1(a). Panel (b): The r-band magnitude for different viewing angles. The line styles are the same as in Figure 1(b). Panel (c): The radio ($\nu=5\,\rm GHz$) light curves for different viewing angles. The line styles are the same as in Figure 1(c). Panel (d): The spectrum evolution for the viewing angle $\theta_{\rm v}=4\theta_c$ case. The line styles are the same as in Figure 1(d).
\label{Figure 2}}
\end{figure}

\begin{figure}
\centering
\subfigure[]{
\begin{minipage}[b]{0.45\textwidth}
\includegraphics[width=1\textwidth]{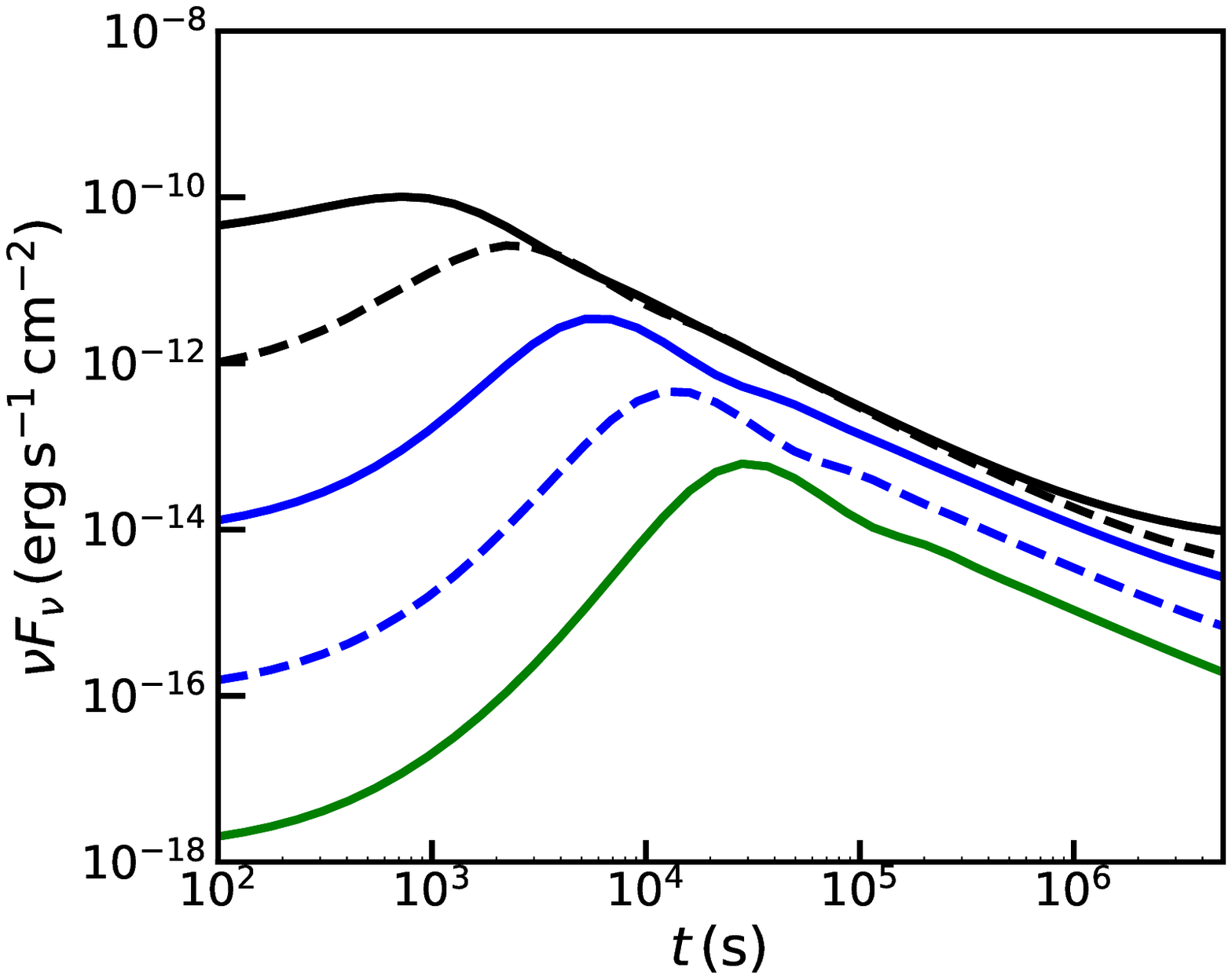}
\label{fig3:subfig:a}
\end{minipage}
}
\subfigure[]{
\begin{minipage}[b]{0.45\textwidth}
\includegraphics[width=1\textwidth]{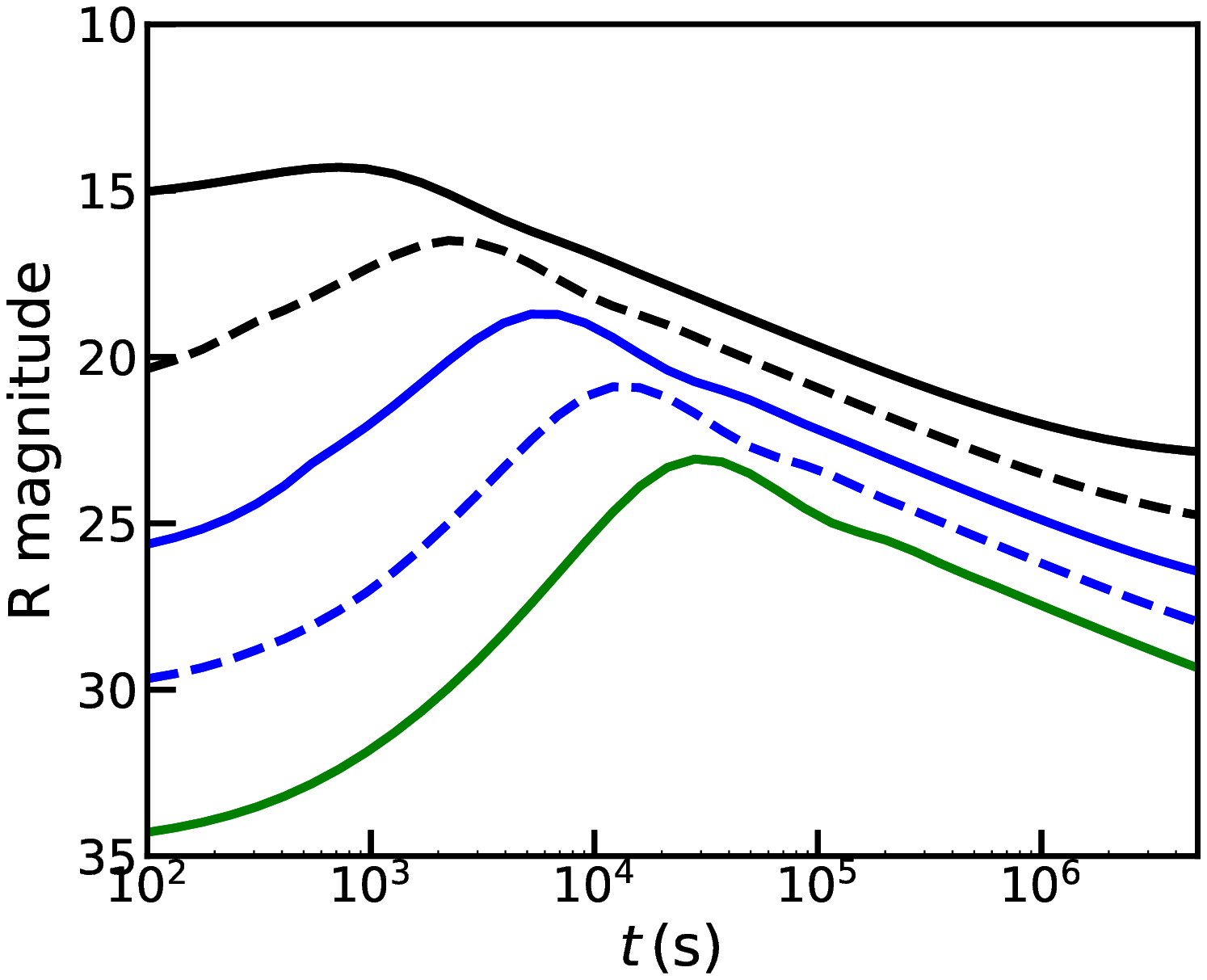}
\label{fig3:subfig:b}
\end{minipage}
}
\subfigure[]{
\begin{minipage}[b]{0.45\textwidth}
\includegraphics[width=1\textwidth]{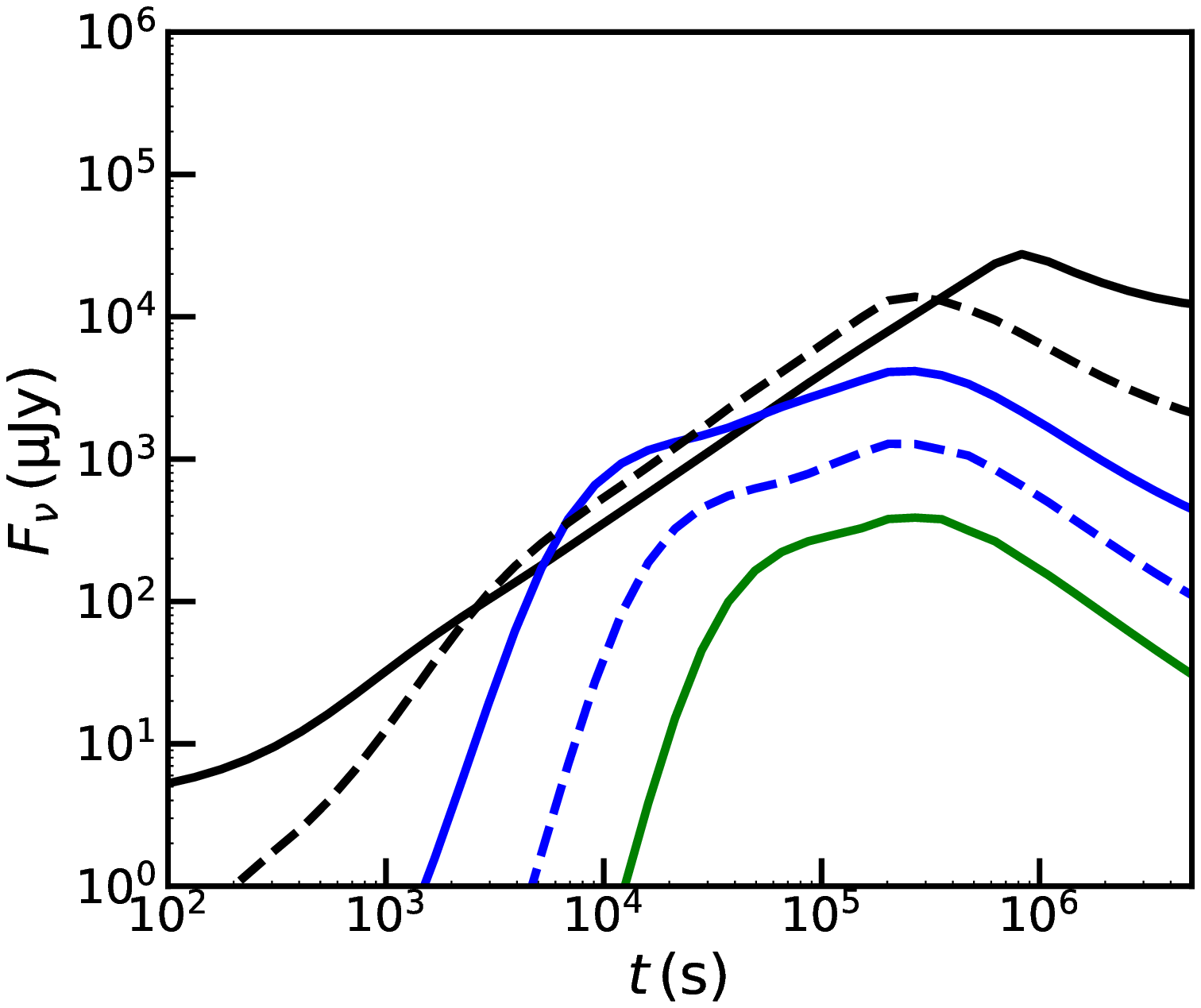}
\label{fig3:subfig:c}
\end{minipage}
}
\subfigure[]{
\begin{minipage}[b]{0.45\textwidth}
\includegraphics[width=1\textwidth]{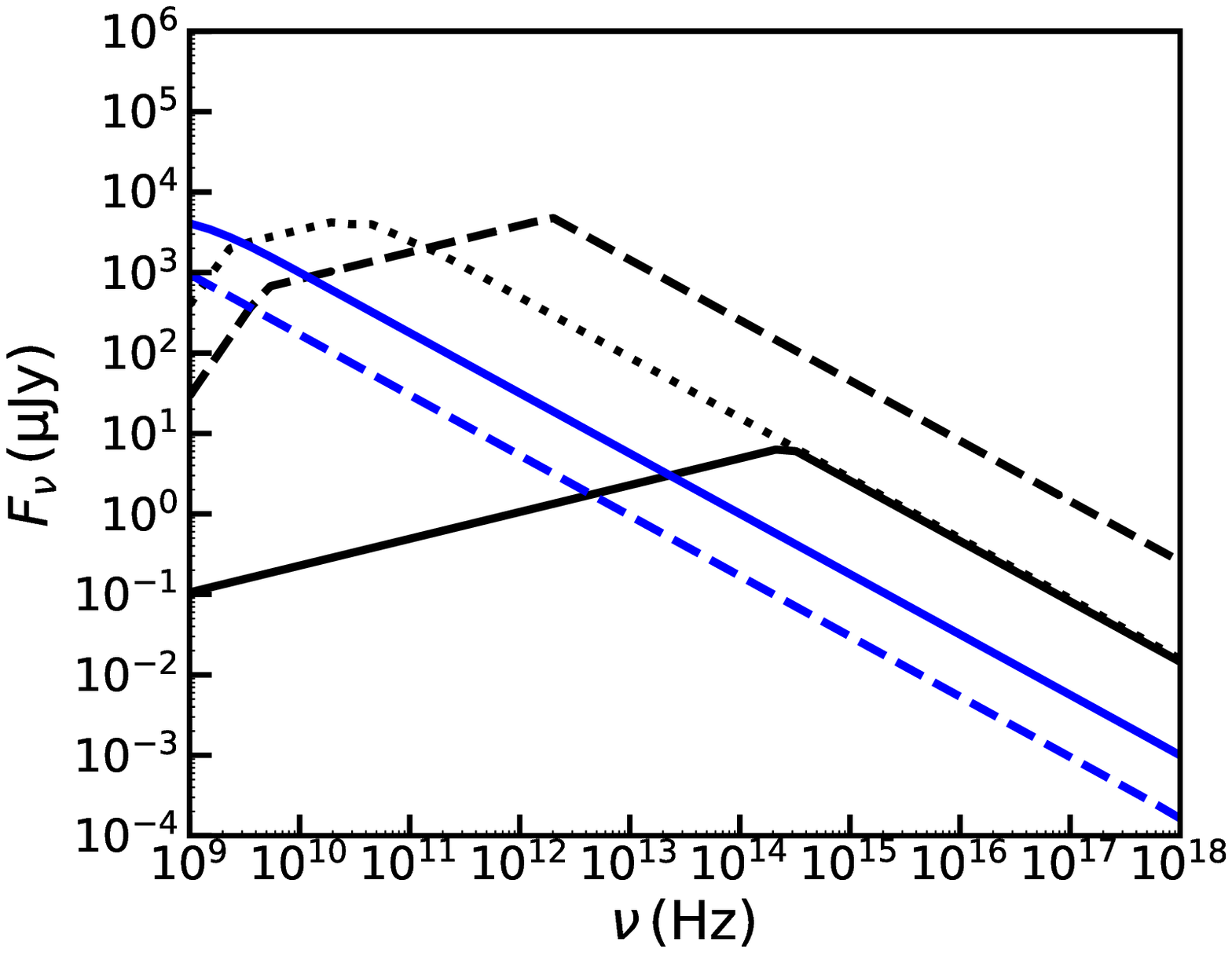}
\label{fig3:subfig:d}
\end{minipage}
}
\caption{The theoretical results in the quasi-isotropic fireball model. Panel (a): The X-ray light curves for different medium densities. The black solid, dashed, blue solid, dashed and green solid lines represent $n=1,\,10^{-1},\,10^{-2},\,10^{-3},\,10^{-4}\,\rm cm^{-3}$ respectively. Panel (b): The r-band magnitude for different medium densities. The line styles are the same as in Figure 3(a). Panel (c): The radio ($\nu=5\,\rm GHz$) light curves for different medium densities. The line styles are the same as in Figure 3(a). Panel (d): The spectrum evolution for $n=0.01\,\rm cm^{-3}$ case. The line styles are the same as in Figure 1(d).
\label{Figure 3}}
\end{figure}

\begin{figure}
\centering
\subfigure[]{
\begin{minipage}[b]{0.45\textwidth}
\includegraphics[width=1\textwidth]{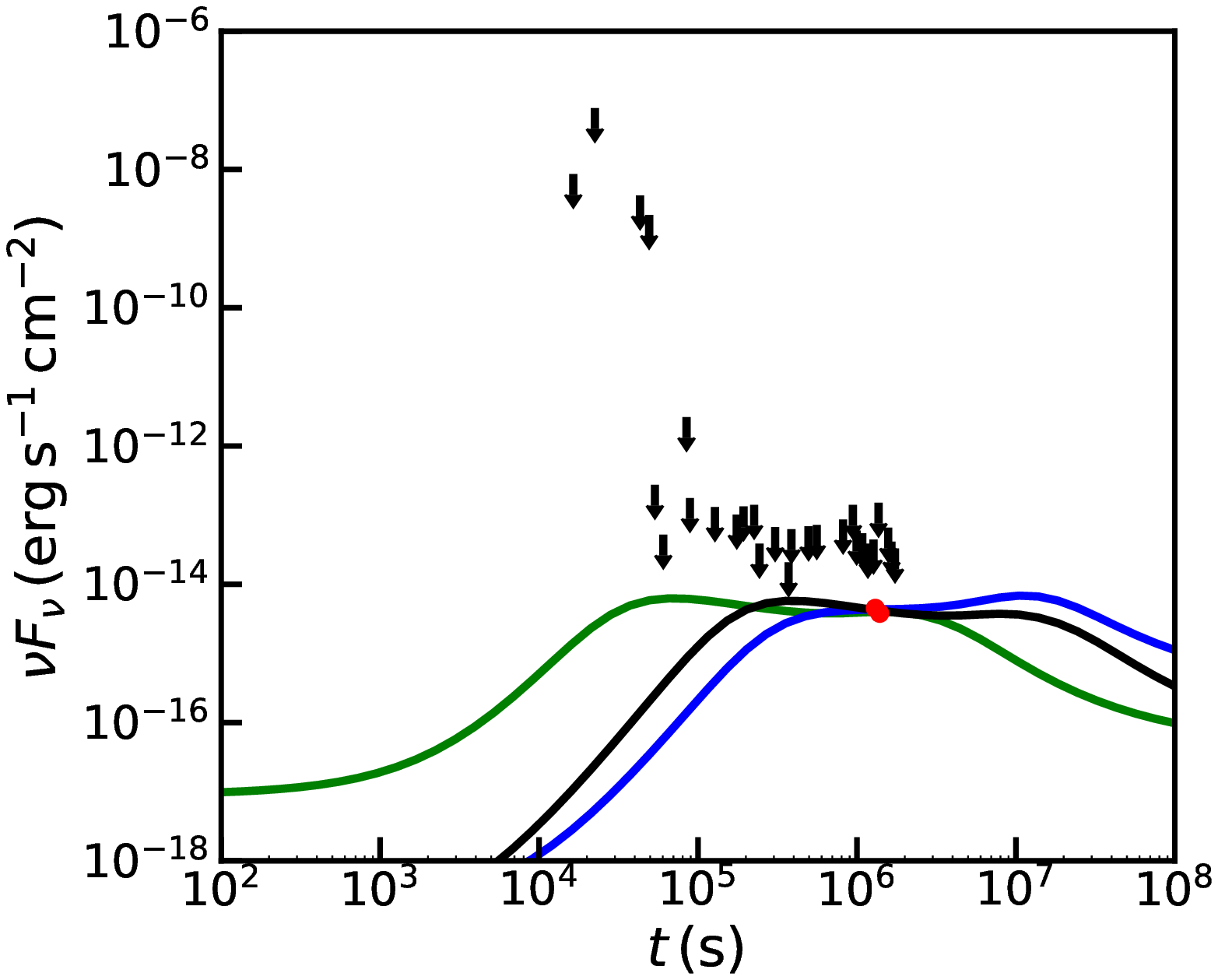}
\label{fig4:subfig:a}
\end{minipage}
}
\subfigure[]{
\begin{minipage}[b]{0.45\textwidth}
\includegraphics[width=1\textwidth]{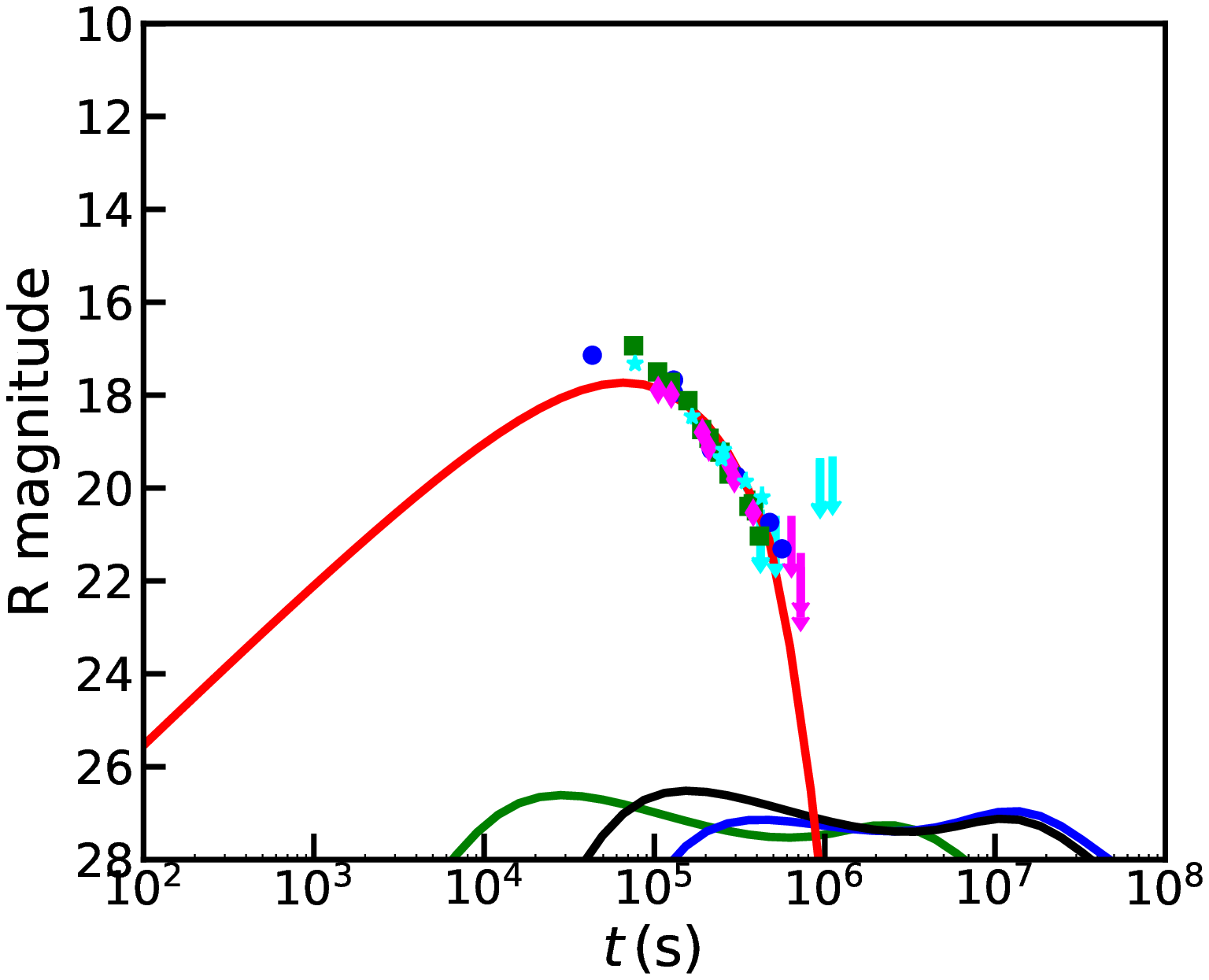}
\label{fig4:subfig:b}
\end{minipage}
}
\subfigure[]{
\begin{minipage}[b]{0.45\textwidth}
\includegraphics[width=1\textwidth]{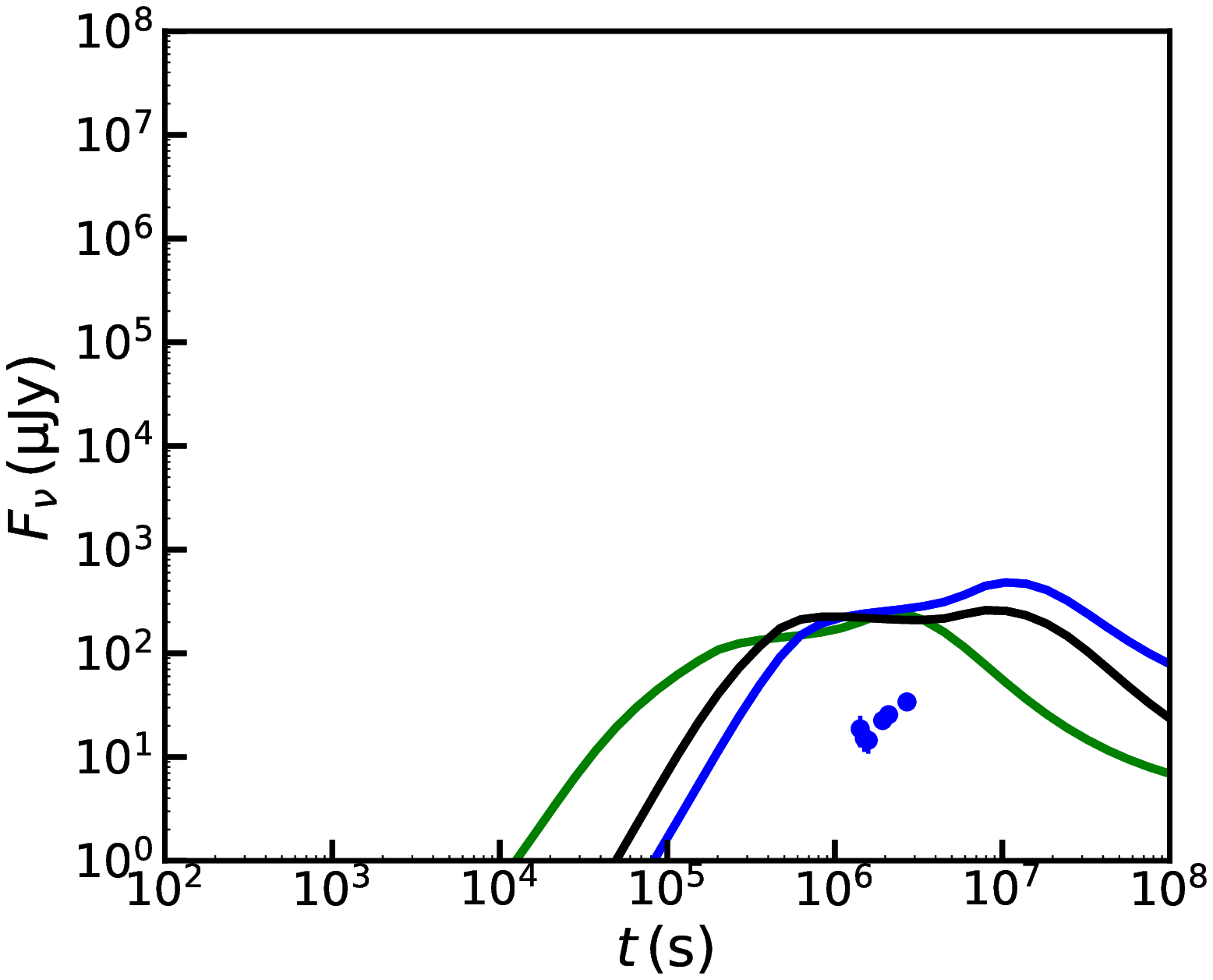}
\label{fig4:subfig:c}
\end{minipage}
}
\caption{Fitting multi-wavelength data in the structured jet model, with parameters given in Table 1. Panel (a): The fitting of X-ray data (including black upper limits and red data-points) \citep{swinupaper17, chan17} with the structured jet model. Fitting parameters for different lines are shown in Table 1. Panel (b): The fitting of r-band data (blue datapoints-\citet{pian17}; green datapoints-\citet{arc17}; magenta datapoints and upper limits-\citet{sma17}; cyan datapoints and upper limits-\citet{and17}) with the kilonova emission plus the structured jet component. Panel (c): The comparison of the predicted flux with radio observations (blue datapoints) \citep{hall17}.
\label{Figure 4}}
\end{figure}

\begin{figure}
\centering
\subfigure[]{
\begin{minipage}[b]{0.45\textwidth}
\includegraphics[width=1\textwidth]{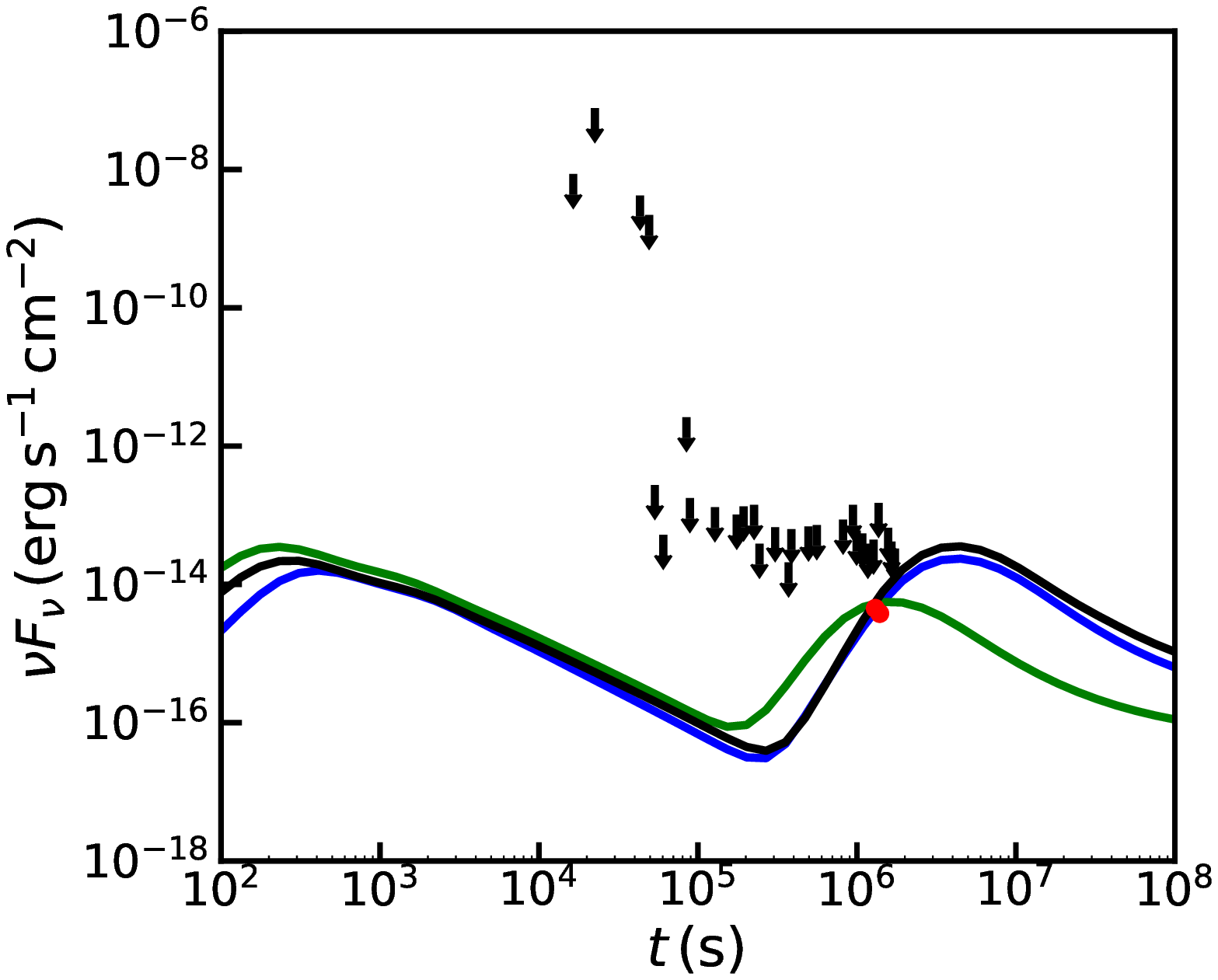}
\label{fig5:subfig:a}
\end{minipage}
}
\subfigure[]{
\begin{minipage}[b]{0.45\textwidth}
\includegraphics[width=1\textwidth]{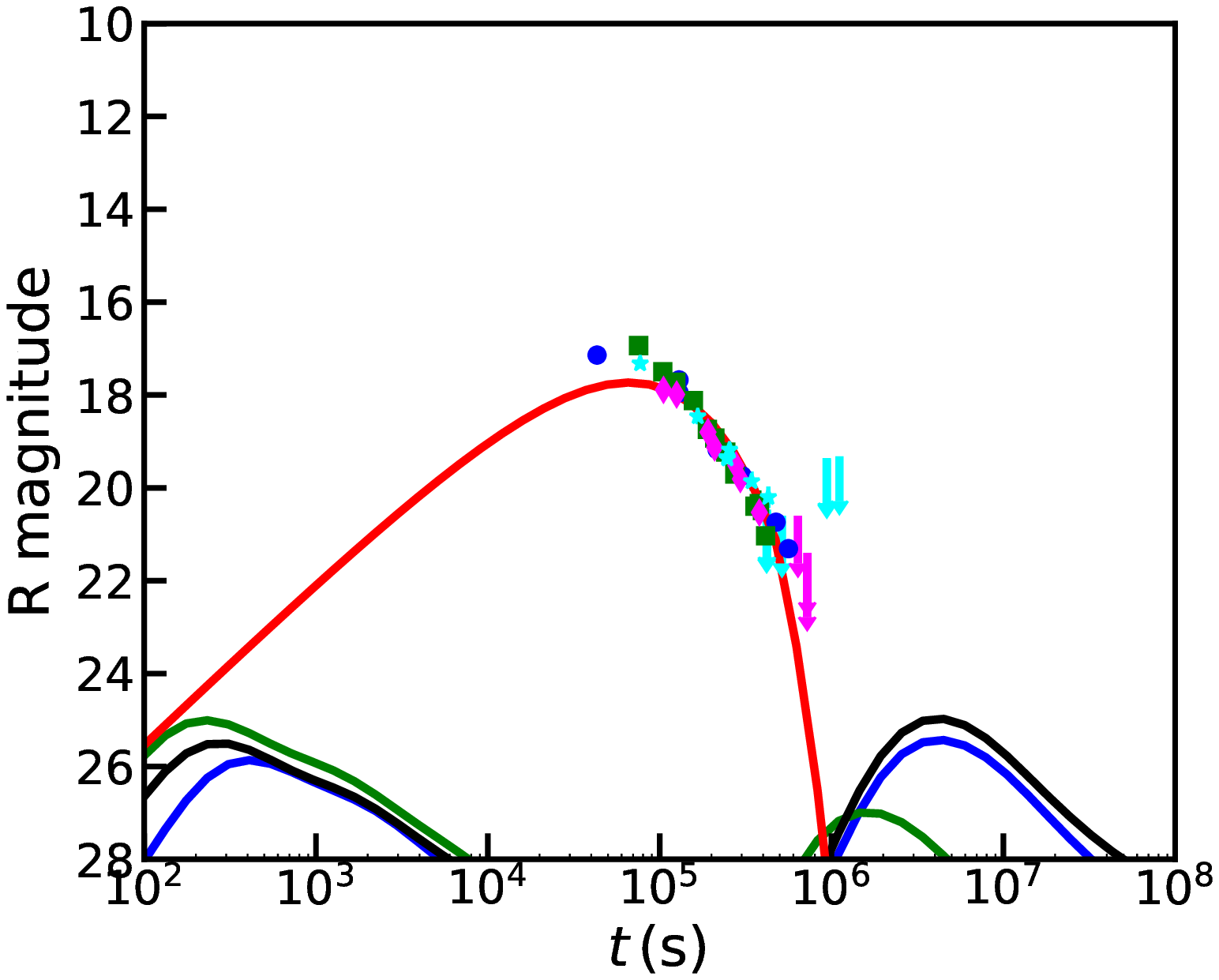}
\label{fig5:subfig:b}
\end{minipage}
}
\subfigure[]{
\begin{minipage}[b]{0.45\textwidth}
\includegraphics[width=1\textwidth]{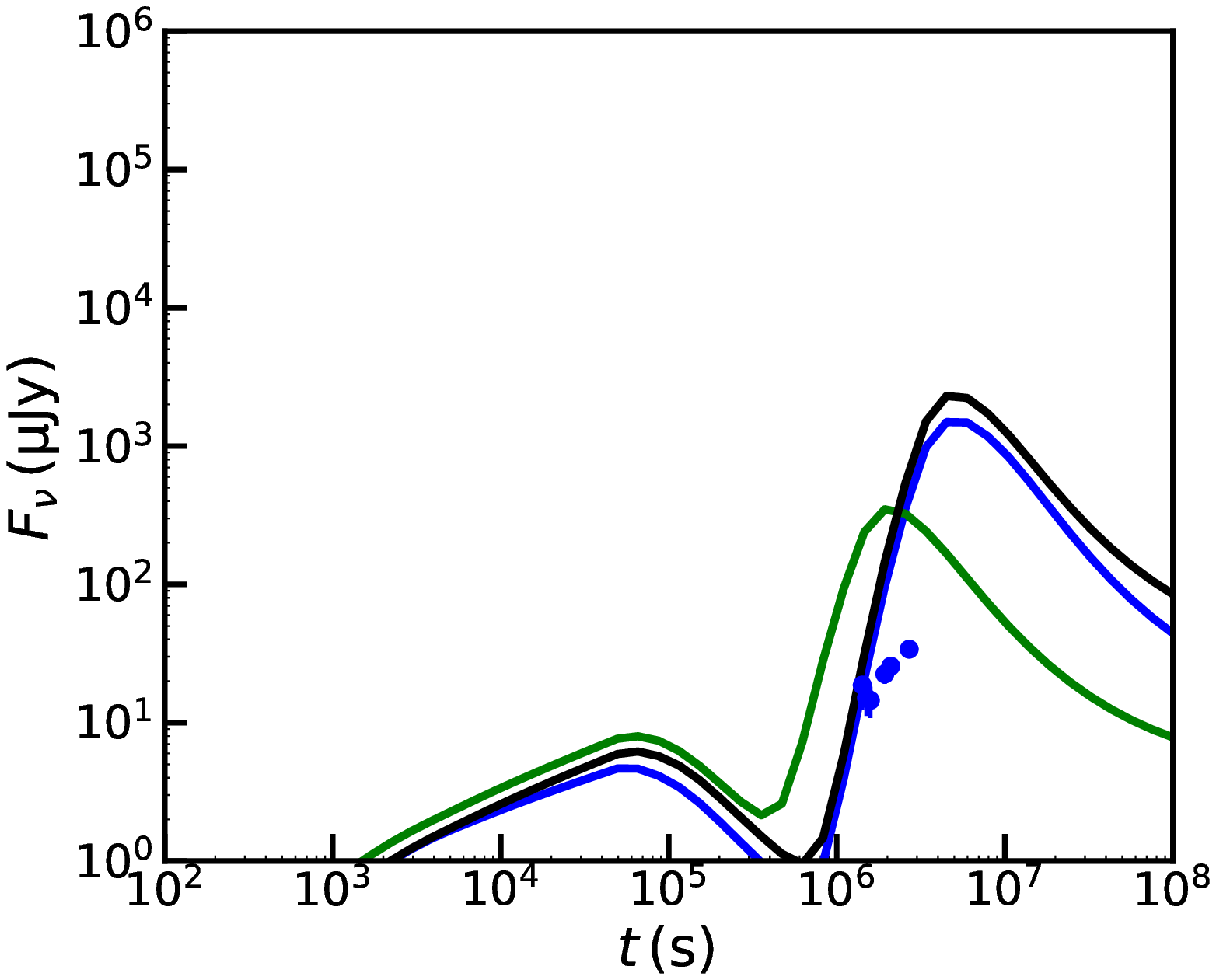}
\label{fig5:subfig:c}
\end{minipage}
}
\caption{Fitting of multi-wavelength data in the two-component jet model, with parameters given in Table 2. Panel (a): The fitting of X-ray data (including black upper limits and red data-points) \citep{swinupaper17, chan17} with the two-component jet model. Fitting parameters for different lines are shown in Table 2. Panel (b): The fitting of r-band data (blue datapoints-\citet{pian17}; green datapoints-\citet{arc17}; magenta datapoints and upper limits-\citet{sma17}; cyan datapoints and upper limits-\citet{and17}) with the kilonova emission plus the two-component jet component. Panel (c): The comparison of the predicted flux with radio observations (blue datapoints) \citep{hall17}.
\label{Figure 5}}
\end{figure}

\begin{figure}
\centering
\includegraphics[width=0.6\textwidth]{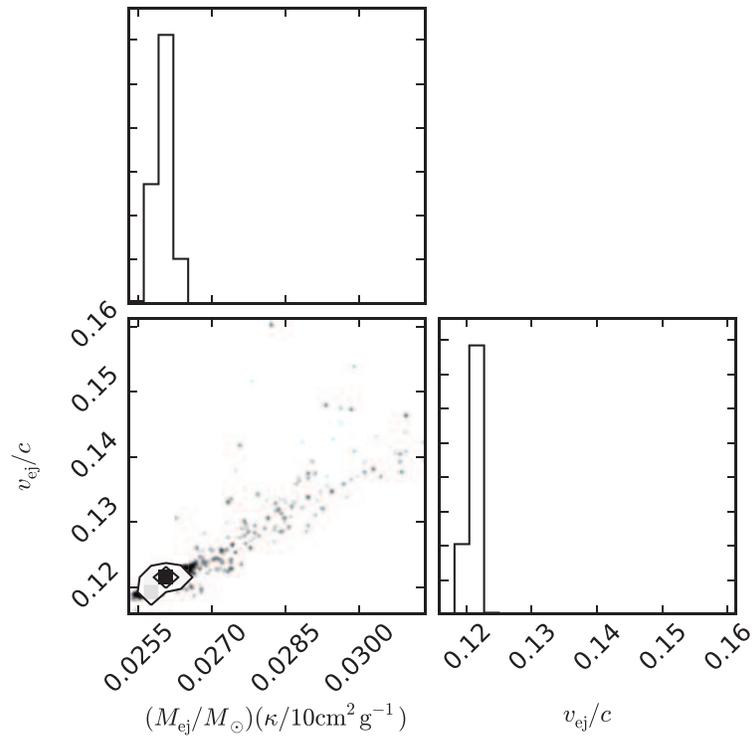}
\caption{Parameter corner in modeling of r-band data. The contours are 1$\sigma$, 2$\sigma$, and 3$\sigma$ uncertainties,
respectively.
\label{Figure 6}}
\end{figure}

\clearpage

\begin{table}
\begin{center}
\caption{Fitting parameters of Figure 4.}
\begin{tabular}{ccccc}
\hline
\hline
& $\varepsilon_0$(erg/sterad) & $\theta_{\rm v}$ & $n(\rm cm^{-3})$& $\Gamma_0$ \\
\hline
Black line & $10^{50}$  & $5\theta_c$ & $5\times10^{-4}$ &300  \\
\hline
Green line & $10^{49}$  & $5\theta_c$ & $ 6\times10^{-3}$ & 300 \\
\hline
Blue line & $10^{50}$  & $5.5\theta_c$ & $ 10^{-3}$ & 300 \\
\hline
\end{tabular}
\end{center}

\end{table}

\begin{table}
\centering
\caption{Fitting parameters of Figure 5.}
\begin{tabular}{ccccccc}
\hline
\hline
& $\varepsilon_{\rm out}$(erg/sterad) &$\varepsilon_{\rm in}$(erg/sterad) & $\theta_{\rm v}$ & $n(\rm cm^{-3})$ & $\Gamma_{\rm in}$ & $\Gamma_{\rm out}$ \\
\hline
Black line & $10^{46}$  & $10^{50}$ & $4\theta_c$ & $3\times 10^{-3}$ & 300 & 30 \\
\hline
Green line &  $10^{46}$  & $10^{49}$ & $4\theta_c$ & $5\times 10^{-3}$ & 300 & 30\\
\hline
Blue line & $10^{46}$ & $10^{50}$ & $3.6\theta_c$ & $10^{-3}$ & 300 & 30 \\
\hline
\end{tabular}

\end{table}


\begin{thebibliography}{}

\bibitem[Abadie et al.(2010)]{aba10}Abadie, J., Abbott, B. P., Abbott, R., et al. 2010, Classical and Quantum Gravity, 27, 173001
\bibitem[Abbott et al.(2016a)]{abb16a}Abbott, B. P., Abbott, R., Abbott, T. D., et al. 2016a, \prl, 116, 241103
\bibitem[Abbott et al.(2016b)]{abb16b}Abbott, B. P., Abbott, R., Abbott, T. D., et al. 2016b, \prl, 116, 061102
\bibitem[Abbott et al.(2017a)]{abb17a}Abbott, B. P., Abbott, R., Abbott, T. D., et al. 2017a, \prl, 118, 221101
\bibitem[Abbott et al.(2017b)]{abb17b}Abbott, B. P., Abbott, R., Abbott, T. D., et al. 2017b, \prl, 119, 141101
\bibitem[Abbott et al.(2017c)]{abb17c}Abbott, B. P., Abbott, R., Abbott, T. D., et al. 2017c, \prl, 119, 161101
\bibitem[Abbott et al.(2017d)]{abb17d}Abbott, B. P., Abbott, R., Abbott, T. D., et al. 2017d, \apjl, 848, L12
\bibitem[Andreoni et al.(2017)]{and17} Andreoni, L., Ackley, K., Cooke, J., et al.\ 2017, PASA, in press % r-band data Aus paper
\bibitem[Arcavi et al.(2017)]{arc17}Arcavi, I., Hosseinzadeh, G., Howell, D., et al. 2017, \nat, 551, 64
\bibitem[Barnes \& Kasen(2013)]{bar13} Barnes, J., \& Kasen, D.\ 2013, \apj, 775, 18
%\bibitem[Barnes et al.(2016)]{bar16} Barnes, J., Kasen, D., Wu, M.-R., \& Mart{\'{\i}}nez-Pinedo, G.\ 2016, \apj, 829, 110
\bibitem[Berger(2014)]{ber14}Berger, E. 2014, \araa, 52, 43
\bibitem[Blandford \& McKee(1976)]{bm76}Blandford, R. D., \& McKee, C. F. 1976, Phys. Fluids, 19, 1130
\bibitem[Bogomazov et al.(2007)]{bog07}Bogomazov, A. I., Lipunov, V. M., \& Tutukov, A. V. 2007, ARep, 51, 308
\bibitem[Connaughton et al.(2016)]{con16}Connaughton, V., Burns, E., Goldstein, A., et al. 2016, \apjl, 826, L6
\bibitem[Coulter et al.(2017)]{cou17}Coulter, D. A., Foley, R. J., Kilpatrick, C. D., et al. 2017, Science, doi:10.1126/science.aap9811, (arXiv:1710.05452)
%\bibitem[Connaughton et al.(2017)]{con17}Connaughton, V., Blackburn, L., Briggs, M. S., et al. 2017, GCN, 21506
%\bibitem[Dai(2004)]{dai2004} Dai, Z. G. 2004, \apj, 606, 1000
\bibitem[Dai \& Gou(2001)]{dai01}Dai, Z. G., \& Gou, L. J. 2001, \apj, 552, 72
%\bibitem[Dai \& Lu(1998a)]{dai1998a} Dai, Z. G., \& Lu, T. 1998a, \aap, 333, L87
%\bibitem[Dai \& Lu(1998b)]{dai1998b} Dai, Z. G., \& Lu, T. 1998b, \prl, 81, 4301
\bibitem[Dai et al.(2006)]{dai2006} Dai, Z. G., Wang, X. Y., Wu, X. F., \& Zhang, B. 2006, Science, 311, 1127
\bibitem[de Mink \& King(2017)]{demi17}de Mink, S. E., \& King, A. 2017, \apjl, 839, L7
\bibitem[Dietrich \& Ujevic(2017)]{die17} Dietrich, T., \& Ujevic, M.\ 2017, Classical and Quantum Gravity, 34, 105014
\bibitem[Drout et al.(2017)]{dro17}Drout, M. R., Piro, A. L., Shappee, B. J., et al. 2017, Science, doi:10.1126/science.aaq0049, (arXiv:1710.05443)
\bibitem[Eichler et al.(1989)]{eich89}Eichler, D., Livio, M., Piran, T., \& Schramm, D. N. 1989, \nat, 340, 126
\bibitem[Evans et al.(2017)]{swinupaper17} Evans, P., Cenko, S., Kennea, J. A., et al. 2017, Science, doi:10.1126/science.aap9580 (arXiv:1710.05437)%Swift/NuSTAR observation
\bibitem[Faber et al.(2006)]{fab06}Faber J. A., Baumgarte T. W., Shapiro S. L., Taniguchi K., 2006, \apjl, 641, L93
\bibitem[Fong et al.(2015)]{fong15}Fong, W., Berger, E., Margutti, R., \& Zauderer, B. A. 2015, \apj, 815, 102
\bibitem[Ghirlanda et al.(2016)]{ghir16}Ghirlanda, G., Salafia, O. S., Pescalli, A., et al. 2016, \aap, 594, 84
\bibitem[Giacomazzo et al.(2013)]{gia13}Giacomazzo B., Perna R., Rezzolla L., Troja E., Lazzati D., 2013, \apjl, 762, L18
\bibitem[Goldstein et al. (2017)]{fermi17}Goldstein, A., Veres, P., Burns, E. et al., 2017, \apjl, 848, L14   %fermi detection paper
\bibitem[Gottlieb et al.(2017a)]{got17}Gottlieb, O., Nakar, E., \& Piran, T. 2017a, arXiv:1705.10797
\bibitem[Gottlieb et al.(2017b)]{got17b}Gottlieb, O., Nakar, E., Piran, T., \& Hotokezaka, K. 2017b, arXiv:1705.10797
\bibitem[Granot et al.(2002)]{gra02}Granot, J., Panaitescu, A., Kumar, P., \& Woosley, S. E. 2002, \apjl, 570, L61
\bibitem[Hallinan et al.(2017)]{hall17}Hallinan, G., Corsi, A., Mooley, K. P., et al. 2017, Science, doi:10.1126/science.aap9855 (arXiv:1710.05435)
\bibitem[Hjorth et al.(2017)]{hjo17}Hjorth, J., Levan, A. J., Tanvir, N. R. 2017, \apjl, 848, L31
\bibitem[Hotokezaka \& Piran(2015)]{hot15}Hotokezaka, K., \& Piran, T. 2015, \mnras, 450, 1430
\bibitem[Huang et al.(1999)]{huang99}Huang, Y. F., Dai, Z. G., \& Lu, T. 1999, \mnras, 309, 513
\bibitem[Huang et al.(2000)]{huang00}Huang, Y. F., Gou, L. J., Dai, Z. G., \& Lu, T. 2000, \apj, 543, 90
\bibitem[Huang et al.(2004)]{huang04}Huang, Y. F., Wu, X. F., Dai, Z. G., Ma, H. T., \& Lu, T. 2004, \apj, 605, 300
\bibitem[Jin et al.(2017)]{jin17}Jin, Z. P., Li. X., Wang, H., et al. 2017, arXiv:1708.07008
\bibitem[Kasen et al.(2013)]{kas13}Kasen, D., Badnell, N. R., \& Barnes, J. 2013, \apj, 774, 25
\bibitem[Kasliwal et al.(2017)]{kas17}Kasliwal, M. M., Nakar, E., Singer, L. P., et al. 2017, Science, doi:10.1126/science.aap9455 (arXiv:1710.05436)
\bibitem[Kathirgamaraju et al.(2017)]{kath17}Kathirgamaraju, A., Duran, R. B., \& Giannios, D. 2017, arXiv:1708.07488v1
\bibitem[Kawaguchi et al.(2016)]{kaw16} Kawaguchi, K., Kyutoku, K., Shibata, M., \& Tanaka, M.\ 2016, \apj, 825, 52
\bibitem[Kilpatrick et al.(2017)]{kil17}Kilpatrick, C. D., Foley, R. J., Kasen, D., et al. 2017, Science, doi:10.1126/science.aaq0073, (arXiv:1710.05434)
\bibitem[Kisaka et al.(2017)]{kis17}Kisaka, S., Ioka, K., Kashiyama, K., \& Nakamura, T. 2017, arXiv:1711.00243
\bibitem[Korobkin et al.(2012)]{kor12} Korobkin, O., Rosswog, S., Arcones, A., \& Winteler, C.\ 2012, \mnras, 426, 1940
\bibitem[Kulkarni(2005)]{kul05} Kulkarni, S.~R. 2005, arXiv:astro-ph/0510256
\bibitem[Kumar \& Granot(2003)]{kum03}Kumar, P., \& Granot, J. 2003, \apj, 591 ,1075
\bibitem[Kyutoku et al.(2015)]{kyu15} Kyutoku, K., Ioka, K., Okawa, H., Shibata, M., \& Taniguchi, K.\ 2015, \prd, 92, 044028
\bibitem[Kyutoku et al.(2013)]{kyu13} Kyutoku, K., Ioka, K., \& Shibata, M.\ 2013, \prd, 88, 041503
\bibitem[Lamb \& Kobayashi(2017)]{lamb17}Lamb, G. P., \& Kobayashi, S. 2017, arXiv:1706.03000v1
\bibitem[Lazzati et al.(2017a)]{laz17a}Lazzati, D., Deich, A., Morsony, B. J., \& Workman, J. C. 2017, \mnras, 471, 1652
\bibitem[Lazzati et al.(2017b)]{laz17b}Lazzati, D., L\'opez-C\'amara, D., Cantiello, M., et al. 2017, arXiv:1709.01468v2
\bibitem[Li \& Paczy\'nski(1998)]{li98}Li, L. X., \& Paczy\'nski, B. 1998, \apjl, 507, L59
\bibitem[Liang et al.(2010)]{liang10}Liang, E. W., Yi, S. X., Zhang, J., et al. 2010, \apj, 725, 2209
\bibitem[Loeb(2016)]{loeb16}Loeb, A. 2016, \apjl, 819, L21
\bibitem[L\"u et al.(2012)]{lv12}L\"u, J., Zou, Y. C., Lei, W. H., et al. 2012, \apj, 751, 49
\bibitem[Martynov et al.(2016)]{mar16}Martynov, D. V., Hall, E. D., Abbott, B. P., et al. 2016, \prd, 93, 112004
\bibitem[Metzger et al.(2010)]{met10}Metzger, B. D., Martinez-Pinedo, G., Darbha, S., et al. 2010, \mnras, 406, 2650
\bibitem[Metzger \& Berger(2012)]{met12}Metzger, B. D., \& Berger, E. 2012, \apj, 746, 48
\bibitem[Metzger(2017)]{met17} Metzger, B.~D.\ 2017, Living Reviews in Relativity, 20, 3
\bibitem[Mochkovitch et al.(1993)]{moch93}Mochkovitch, R., Hernanz, M., Isern, J., \& Martin, X. 1993, \nat, 361, 236
\bibitem[Moderski et al.(2000)]{mod00}Moderski, R., Sikora, M., \& Bulik, T. 2000, \apj, 529, 151
%\bibitem[Moldon et al. (2017a)]{ragcn17a}  Moldon, J., Beswick, R., et al. 2017a, GCN, 21804, 1   %Radio GCN upper limits
%\bibitem[Moldon et al. (2017b)]{ragcn17b}  Moldon, J., et al. 2017b, GCN, 21940, 1   %Radio GCN upper limits
\bibitem[Murguia-Berthier et al.(2017)]{murg17}Murguia-Berthier, A., Ramirez-Ruiz, E., Kilpatrick, C. D., et al. 2017, ApJL, 848, L34
\bibitem[Nagakura et al.(2014)]{nag14}Nagakura, H., Hotokezaka, K., Sekiguchi, Y., Shibata, M., \& Ioka, K. 2014, \apj, 784, 28
\bibitem[Nakar(2007)]{nak07}Nakar, E. 2007, Phys. Rep., 442, 166
\bibitem[Nakar \& Piran(2011)]{nak11}Nakar, E., \& Piran, T. 2011, \nat, 478, 82
\bibitem[Nakar \& Piran(2017)]{nak17}Nakar, E., \& Piran, T. 2017, \apj, 834, 28
\bibitem[Narayan et al.(1992)]{nar92}Narayan, R., Paczynski, B., \& Piran, T. 1992, \apjl, 395, L83
\bibitem[Panaitescu \& M\'esz\'aros(1998)]{pan98}Panaitescu, A., \& M\'esz\'aros, P. 1998, \apjl, 493, L31
\bibitem[Paczynski(1986)]{pac86}Paczynski, B. 1986, \apjl, 308, L43
\bibitem[Perna et al.(2016)]{per16}Perna, R., Lazzati, D., \& Giacomazzo, B. 2016, \apjl, 821, L18
\bibitem[Pian et al.(2017)]{pian17}Pian, E., D' Avanzo, P., Benetti, S., et al. 2017, \nat, 551, 67
\bibitem[Rossi et al.(2002)]{ros02}Rossi, E., Lazzati, D., \& Rees, M. J. 2002, \mnras, 332, 945
\bibitem[Ruiz et al.(2016)]{ruiz16}Ruiz M., Lang R. N., Paschalidis V., Shapiro S. L., 2016, \apjl, 824, L6
\bibitem[Sari(1998)]{sa98}Sari, R. 1998, \apjl, 494, L49
\bibitem[Sari et al.(1998)]{sari98}Sari, R., Piran, T., Narayan, R. 1998, \apjl, 497, L17
\bibitem[Shappee et al.(2017)]{sha17}Shappee, B. J., Simon, J. D., Drout, M. R., et al. 2017, Science, doi:10.1126/science.aaq0186, (arXiv:1710.05432)
\bibitem[Smartt et al.(2017)]{sma17}Smartt, S., Chen, T. W., Jerkstrand, A., et al. 2017, \nat, 551, 75
\bibitem[Sun et al.(2015)]{sun15}Sun, H., Zhang, B., \& Li, Z. 2015, \apj, 812, 33
\bibitem[Tanaka \& Hotokezaka(2013)]{tana13} Tanaka, M., \& Hotokezaka, K.\ 2013, \apj, 775, 113
\bibitem[Tanaka(2016)]{tana16} Tanaka, M.\ 2016, Advances in Astronomy, 2016, 634197
\bibitem[Tanvir et al.(2013)]{tan13} Tanvir, N.~R., Levan, A.~J., Fruchter, A.~S., et al.\ 2013, \nat, 500, 547
\bibitem[Troja et al.(2017)]{chan17}Troja, E., Piro, L., van Eerten, H. et al., 2017, \nat, 551, 71 %x-ray chandra detection data point
\bibitem[Tutukov \& Yungelson(1992)]{tut92}Tutukov, A. V., \& Yungelson, L. R. 1992, \apj, 386, 197
\bibitem[von Kienlin et al.(2017)]{kie17}von Kienlin, A., Meegan, C., \& Goldstein, A. 2017, GCNC, 21520, 1
\bibitem[Vlahakis et al.(2003)]{vla03}Vlahakis, N., Peng, F., \& K\"onigl, A. 2003, \apjl, 594, L23
\bibitem[Waxman(1997)]{wax97}Waxman, E. 1997, \apjl, 491, L49
\bibitem[Xiao \& Dai(2017)]{xiao17}Xiao, D., \& Dai, Z. G. 2017, \apj, 846, 130
\bibitem[Yamazaki et al.(2002)]{yam02}Yamazaki, R., Ioka, K., \& Nakamura, T. 2002, \apjl, 571, L31
\bibitem[Yamazaki et al.(2003)]{yam03}Yamazaki, R., Yonetoku, D., \& Nakamura, T. 2003, \apjl, 594, L79
\bibitem[Yamazaki et al.(2016)]{yam16}Yamazaki, R., Asano, K., \& Ohira, Y. 2016, PTEP, 2016, 051E01
\bibitem[Yu et al.(2013)]{yu2013}Yu, Y.-W., Zhang, B., \& Gao, H.\ 2013, \apjl, 776, L40
\bibitem[Zhang \& M\'esz\'aros(2002a)]{zhang02}Zhang, B., \& M\'esz\'aros, P. 2002a, \apj, 571, 876
\bibitem[Zhang \& M\'esz\'aros(2002b)]{zhang02b}Zhang, B., \& M\'esz\'aros, P. 2002b, \apj, 581, 1236
\bibitem[Zhang(2016)]{zhang16}Zhang, B. 2016, \apjl, 827, L31
\bibitem[Zhang et al.(2017)]{zhangBB17}Zhang, B. B., Zhang, B., Sun, H. et al. 2017, Nature Astronomy submitted (arXiv: 1710.05851)
\end{thebibliography}
\end{document}